\documentclass[useAMS,usenatbib,usegraphicx]{mn2e}
\def\la{\raise.5ex\hbox{$<$}\kern-.8em\lower 1mm\hbox{$\sim$}}
\def\ma{\raise.5ex\hbox{$>$}\kern-.8em\lower 1mm\hbox{$\sim$}}

\def\msol{M$_{\odot}$ }
\def\kms{$\rm km\, s^{-1}$}
\def\cm3{$\rm cm^{-3}$}
\def\Ts{$\rm T_{*}$~}
\def\Vs{$\rm V_{s}$~}
\def\n0{$\rm n_{0}$}
\def\B0{$\rm B_{0}$}
\def\ne{$\rm n_{e}$~}

\def\Te{$\rm T_{e}$}

\def\erg{$\rm erg\, cm^{-2}\, s^{-1}$}
\def\mum{$\mu$m~}

\def\L12{L$_{12\mu m}$~}
\def\F12{F$_{12\mu m}$~}
\def\agr{a$_{gr}$}
\def\Hb{H$_{\beta}$}

\title{An analysis of infrared  emission spectra  from the regions near the Galactic Center}
\author[M. Contini ]{ Marcella Contini$^{ }$  
\\
$^{ }$School of Physics and Astronomy, Tel Aviv University, Tel Aviv
69978, Israel \\
}

\begin{document}

\date{Accepted: Received ; in original form 2009 month day}

\pagerange{\pageref{firstpage}--\pageref{lastpage}} \pubyear{2003}

\maketitle

\label{firstpage}

\begin{abstract}
We present consistent modelling of  line and continuum IR  spectra in the region  close to the
Galactic center. The models account for the coupled effect of shocks and
photoionization from an external source.
The results show that the  shock velocities range between $\sim$ 65 and 80 \kms,
the pre-shock densities between   1\cm3 in the ISM to  200\cm3 in the filamentary structures.
The pre-shock magnetic field increases from 5. 10$^{-6}$ gauss  in  the surrounding  ISM
to  $\sim$ 8 10$^{-5}$ gauss in the Arched Filaments.
The stellar temperatures are $\sim$ 38000 K in the Quintuplet cluster and $\sim$ 27000 K in the Arches Cluster.
The ionization parameter is relatively low ($<$ 0.01 ) with  the highest values near the clusters, reaching
 a maximum  $>$0.01 near the Arches Cluster.
Depletion from the gaseous phase of Si is  found throughout the  whole observed region, indicating the presence of
silicate dust. Grains including iron,  are concentrated throughout  the Arched Filaments.
The  modelling of the continuum SED in the IR range,  indicates that a component
of dust  at temperatures of $\sim$ 100-200 K  is present in the central region of the Galaxy. 
Radio emission appears to be  thermal bremsstrahlung in the E2-W1 filaments  crossing strip,
however a synchrotron component is not excluded. More data are  necessary to resolve this questions.

\end{abstract}


\section{Introduction}

\begin{figure}
\includegraphics[width=0.5\textwidth]{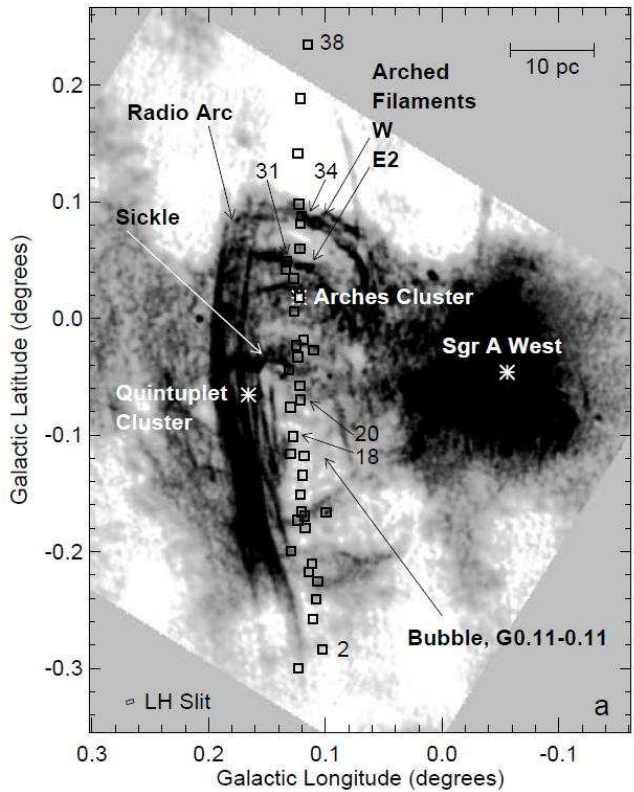}
\caption{Images of the Galactic Center with the observed positions indicated by
square boxes upon the radio continuum (log-log scale) imaged at 21 cm imaged by Yusef-Zadeh
\& Morris (1987)  with 11" resolution (adapted   from Simpson et al.2007, Fig. 1. )
}
\end{figure}

The Galaxy central region cannot be observed  in the optical and UV range
 because of strong extinction (Erickson et al 1991), but in  recent years,
 infrared observations  allowed  a detailed investigation of the gas and dust
structures.

Fig. 1 shows a radio image of this region (Yusef-Zadeh \& Morris 1987). The 
 Sgr A West HII region  contains a quiescent black hole $\sim$ 4 10$^{6}$ \msol (Ghez et al.
 2005; Eisenhauer et al. 2005) which is coincident with the radio source Sgr A* and is located
at the center of the Galaxy. It also contains  a cluster of massive stars.
A distance of 8 kpc is assumed by Simpson et al (2007).
Two  other  clusters of  young massive stars and massive molecular clouds
(Sch\"{o}del et al. 2006) appear in the Galactic Center (GC).  The Arches Cluster and the Quintuplet Cluster
are located $\sim$  25 pc away in the plane of the sky.
The Arches Cluster (Nagata et al 1995 and Cotera et al. 1996)  is a very massive and dense
cluster of young stars  heating and ionizing the  region of the
Arched Filaments and the Sickle. These  are thermal radio  emitters (e.g. Yusef-Zadeh \& Morris 1987,
Morris \& Yusef-Zadeh 1989, Lang et al. 1997, 2001), while the Radio Arc (Yusef-Zadeh et al 1984)
consists of non-thermally emitting linear filaments perpendicular to the Galactic  plane. 
The stars of the Quintuplet Cluster ionize the clouds in the extended region  including  the Bubble. 
A detailed description of the GC is given by Simpson et al (2007).

Excitation of the gas   is  ascribed to photoionization
 because it was found that the excitation variations   depend on  the  projected distances
from the clusters (Erickson 1991).

The radial velocity field is very complicated. The gas velocity   range
in the Sickle is $\sim$ 40-140 \kms (Yusef-Zadeh et al 1997, Lang et al. 1997),
and seems lower closest to Sgr A. Interestingly, in both the Arched Filaments and the Sickle
the velocity of the molecular clouds is similar to that of the gas.

According to the morphology in Fig. 1, it  appears   that the cloud structures 
  characterised by a system of semicircular arcs, 
 arise from stellar winds or supernova explosions. 
This can be noticed for instance,  in the Bubble region (Levine et al. 1999)
  and in the “Radio Arc Bubble” (Rodr\'{i}guez-Fern\'{a}ndez et al. 2001). 
At the same position on the plane of the sky as the Bubble, there is a dense molecular 
cloud,  G0.011−0.011 (Tsuboi et al. 1997) or G0.013−0.013 (Oka et al. 2001). 

Stellar winds and supernova explosions  suggest that the shocks  have a non-negligible 
role in  heating and ionizing both the gas  and the dust.
The fragmented filamentary  structures  characteristic of the GC strengthens this hypothesis.

 The Arches Cluster has also been  investigated in the light of
dynamical evolution of compact young clusters (e.g. Kim, Morris, \& Lee 1999, Kim et al. 2000).

 In the X-ray domain,
using the Chandra telescope, Yusef-Zadeh et al. (2002)  detected three X-ray
components  associated with the  Arches Cluster. 
They claim that hot (10$^7$K)  X-ray emitting gas
is produced by an interaction  of material expelled by the massive stellar winds with
the local interstellar medium.
 One of the sources  has roughly the characteristics expected from
shock-heated gas created by the collision of  a number of 1000 \kms stellar
winds emanating from the stars in the rich dense cluster.
However, the X-ray sources are extended  and hardly related
 to single X-ray binary systems.

Far-infrared (FIR) lines were 
observed using the Kuiper Airborne Observatory (KAO: Colgan et al. 1996; Simpson et al. 1997) 
and the Infrared Space Observatory (ISO: Rodr\'{i}guez-Fern\'{a}ndez et al. 2001; Cotera et al. 2005). 
For both the Arched Filaments (Colgan et al. 1996; Cotera et al. 2005) and the Sickle 
(Simpson et al. 1997; Rodr\'{i}guez-Fern\'{a}ndez et al. 2001), the excitation decreases with 
distance from the Arches Cluster and Quintuplet Cluster, respectively, as 
expected for photoionization.

 Spitzer observations  of MIR spectra
in  38 positions along a line approximately perpendicular to the Galactic plane in the GC
are presented by Simpson et al (2007),
 who    analyse the sources of excitation of the Arched Filaments and of the
other thermal arcs.  They are particularly interested  in the Bubble  physical conditions
relatively to the clusters and the other filament structures.
The observed  positions  are shown in Fig. 1.
Their spectra contain high and low ionization level lines (e.g. [OIV], [SIV], [NeIII], [NeII],
[SIII], [FeIII], [FeII], and [SiII]).

In their paper,  Simpson et al. (2007) use the new observations to determine the stellar properties 
of the most massive stars in the Arches Cluster. 
However,  the  modelling of the spectral line ratios by  pure photoionization codes, (e.g. CLOUDY etc.) 
was  successful only to explain  some line ratios in a few positions.  Simpson et al. (2007) 
conclude that  the  models accounting for the shocks by Contini \& Viegas (2001) 
could explain the relatively strong  [OIV] lines. This  induced us
 to  revisit  Simpson et al.   observations
of  the  Galactic  central region,  by a detailed modelling of the line and continuum spectra, 
constraining the results by  the  interpretation of
the spectra  previously  observed by Erickson et al. (1991) and Colgan et al.(1996).
We adopt for the calculations the code SUMA which accounts for both photoionization and shocks.
In particular, all the lines available in each spectrum and the continuum spectral energy distribution
(SED)  will be modelled in a   consistent way.
The  results will  explain the excitation
 mechanisms  of the gas  near the GC  as well as some particular features, 
e.g. the distribution of dust.
The calculation details are described in Sect. 2. 
 In Sect. 3 the spectra presented by Simpson et al. are modelled and discussed.
In Sect. 4,  line and continuum spectra are modelled
for  Position C - G0.095+0.012  and the E2 thermal radio Filament which were observed by Erickson et al (1991).
In Sect. 5 we examine the observations of Colgan et al (1996)  in the different positions   crossing the
 E2-W1 arched radio filament.
Concluding remarks follow in Sect. 6.

\section{The modelling method}

The  physical parameters are combined  throughout the calculation  of
forbidden and permitted
lines (see Osterbrock 1988) emitted from a shocked nebula. Besides the element abundances,
the main parameters are known to be :
the electron density \ne, the electron temperature \Te,
the component of the magnetic field perpendicular to the flow direction $B$,
the flux from the external source,  which is characterised by its spectral type (e.g. a black body)
and intensity (e.g.  the ionization parameter U), and the fractional
abundance of the ions.
They follow the recombination trend of the gas downstream. Therefore, the
precision of the calculations  requires   to divide the downstream region in
many  different slabs corresponding to  the different physical conditions. 
The line and continuum intensities  in the X-ray  - radio range,  are calculated in each 
slab and  integrated throughout the nebula geometrical thickness.

In pure photoionization models
the density  n   is  constant throughout the nebula, while
in a shock wave regime,   
 n  is calculated downstream by the compression equation in each of the single slabs.
Compression depends on n, $B$, and the shock velocity V.

The ranges of the physical conditions in  the  nebula  are deduced,  as a first guess,
 from the observed lines
(e.g. the shock velocity from the FWHM)  and from  the characteristic line ratios (e.g.
 \ne and  \Te).

The observations indicate that a steady state situation can be applied (Cox 1972).
In this case, the time t required for a parcel of gas to cross
the cooling region from the shock front to the recombination zone, for shock waves
with v=100 \kms, is found to be about 1000/\ne years (calculated by the recombination coefficients)
so, for an electron density \ne = 100 \cm3,  t = 10 years.
Shock velocities within the  GC are not likely to change appreciably in that short a time,
so the steady state calculation should be adequate.

In this paper, the
line and continuum spectra emitted by the gas downstream  of the  shock front, 
 are calculated by
SUMA (see http://wise-obs.tau.ac.il/$\sim$marcel/suma/index.htm for
a detailed description).
 The  code  simulates the physical conditions in  an
emitting gaseous cloud
under the coupled effect of photoionisation from an external radiation
source and shocks. The line and continuum emission from the gas
is calculated consistently with dust reprocessed radiation
 in a plane-parallel geometry.

In a composite (shock and photoionization) code such as SUMA, 
the input parameters are: the  shock velocity \Vs, the   preshock density \n0,
the preshock magnetic field \B0 which refer to the shock, while, the colour  temperature of the hot star \Ts
and the ionization parameter $U$ (the number of photons per  number  of electrons at the nebula) refer to the flux. 
The geometrical thickness of the emitting nebula $D$,
the dust-to-gas ratio $d/g$, and the  abundances of He, C, N, O, Ne, Mg, Si, S, A, and Fe relative to H
are also considered.
 The distribution of the grain radius (\agr) downstream
is determined by sputtering,  beginning with an initial  radius. 

The calculations start at the shock front where the gas is compressed
and thermalized adiabatically, reaching the maximum temperature
in the immediate post-shock region
(T$\sim$ 1.5 10$^5$ ($V_s$/100 \kms)$^2$). T decreases downstream
following recombination. The cooling rate is calculated in each slab.
The downstream region is cut up into a maximum of 300  plane-parallel slabs 
with different geometrical widths calculated automatically,  in order
to account for the temperature gradient  (Contini 1997 and references therein).

 In each slab, compression is
calculated by  the Rankine-Hugoniot equations  for the
conservation of mass, momentum and energy throughout the shock front.
Compression (n/\n0) downstream  ranges between
4 (the adiabatic jump) and $\leq$ 10,  depending on  \Vs and \B0.
The stronger the  magnetic field the lower is  compression downstream, while
 a higher   shock velocity  corresponds to a  higher  compression.

The ionizing radiation from an external source is characterized by its
spectrum  depending on \Ts,  and by the ionization parameter. The flux is calculated at 440 energies, 
from a few eV to KeV.
 Due to  radiative transfer, the
spectrum changes throughout the downstream slabs, each of them
contributing to the optical depth. 
In addition to the radiation from the primary
source, the effect of the diffuse radiation created by the gas line and  continuum emission
 is also taken into account,
 using  240 energies to calculate the spectrum.

For each slab of gas, the  fractional abundance of the ions of each
chemical element are obtained by solving the ionization equations.
These equations account for the ionization mechanisms
(photoionization by the primary and diffuse radiation, and
collisional ionization) and recombination mechanisms (radiative, 
dielectronic recombinations), as well as charge transfer effects.
The ionization equations are coupled to the energy equation
when collision processes dominate, and to the thermal balance if
radiative processes dominate. This latter balances the heating
of the gas due to the primary  and diffuse radiations reaching
the slab,  and the cooling, due to recombinations and collisional
excitation of the ions followed by line emission, dust collisional
ionization, and thermal bremsstrahlung. The coupled equations
are solved for each slab, providing the physical conditions necessary
for calculating the slab optical depth, as well as its  line and
continuum emissions. The slab contributions are integrated
throughout the cloud.

In particular, the absolute line fluxes referring to the ionization level i of element K are calculated
by the term n$_K$(i) which represents the density of the ion i. We consider that
n$_K$(i)= X(i) [K/H] N$_H$, where X(i) is the fractional abundance of the
ion i calculated by the ionization equations, [K/H] is the relative abundance of the element K to H,
and N$_H$ is the density of H (by number \cm3). In models  including shock,
N$_H$ is calculated by the compression equation (Cox 1972) in each slab downstream (Sect. 2).
So the  abundances of the elements  are given relative to H as input parameters.

Dust grains are coupled to the gas across the shock front by the magnetic field,
and are heated by radiation
 from the stars  and collisionally by the gas to a maximum temperature which is a function
of the shock velocity,
 of the chemical composition, and the radius of the grains, up to the evaporation temperature
(T$_{dust} \geq$ 1500 K).
The grain radius distribution downstream is determined by sputtering, which depends
on the shock velocity and on the density.
Throughout shock fronts and downstream, the grains might be destroyed by sputtering.

Summarizing, very schematically :

\noindent
1) we  adopt an initial  Te ($\sim$ 10$^4$ K) and the input parameters; 

\noindent
2)  calculate the density from the compression equation;

\noindent
3)  calculate the fractional abundances of the ions from each level for each element;

\noindent
4)  calculate line emission, free-free and free-bound emission;

\noindent
5)  recalculate Te by thermal balancing or the enthalpy equation;

\noindent
6)  calculate the optical depth of the slab and the primary and secondary fluxes;

\noindent
7) adopt the parameters found in slab i as initial conditions for slab i+1;

Integrating  the contribution of  the line intensities 
calculated in each slab, we obtain the absolute
fluxes of  each of the lines, calculated at the nebula (the same for bremsstrahlung).

\noindent
8) We then calculate the line ratios to a certain line (in the present case [SIII], because
we do not have values of \Hb or other H lines)

\noindent
9)  and compare them with the observed line ratios.

The observed data have errors, both random and systematic.
 Models are generally allowed to reproduce the data within a factor of 2.
This leads to input parameter   ranges of  a few percent.
The uncertainty in the calculation results, on the other hand, derives from the use of many atomic parameters,
such as recombination coefficients, collision strengths, etc., which are continuously
updated. Moreover, the precision of the integrations  depends on the computer efficiency.

Simpson et al. present the spectra observed in close regions throughout the slit.
Actually, we are interested in the trend of the physical conditions, as well as in the
parameter ranges. Therefore, to avoid the  accumulation of errors which leads to inconsistent
results, we try to reproduce the data as close as possible.
If the calculated lines are compatible with the observed ones within the error
of each line, we adopt the input parameters as the result in this observed position.
If there are discrepancies, we  change consistently the input parameters and we restart the whole
calculation process. As explained in the text, the alterations are done on the basis that
 \Ts  affects [NeIII]/[NeII], U  affects [FeIII]/[FeII], [OIV] depends on  \Vs, etc.
We  also change the relative abundances to obtain a good fit for all the line ratios, however, 
  they affect the cooling rate, so it is important to restart
the calculation process each time.

As a matter of fact, a degeneracy can arise e.g. from the density and the
magnetic field which are directly correlated.
There can be a doubt in the values of \n0 and \B0 leading to the
best fit of the observations in a certain position $p$. Our  method is to adopt
as a first guess, the input parameters  of position $p$-1, and then  modify
them, little by little,  calculating a grid of models,
until all the observed line ratios in position $p$ are satisfactorily reproduced within the
least error.

The number of ionizing photons  cm$^{-2}$ s$^{-1}$ produced by the hot source is
$N$= $\int_{\nu 0}B_{\nu} /(h\nu ) d\nu$, where 
 $\nu$0=3.29 10$^{15}$ s$^{-1}$ and B$_{\nu}$ is the Planck function.
The  flux  from the star   is combined  with U and n
by :  
 $N$ (r/R)$^2$ = U n c, where 
 r is the radius of the hot source (the stars),
R is the radius
of the nebula (in terms of the distance from  the stars), n is the density of the nebula,
and c the speed of light.
Therefore, \Ts and U compensate each other, but only in a qualitative way, because
 \Ts  determines the frequency  distribution of the primary flux, while U
represents the number of photons per number of electrons reaching the nebula.
The choice of \Ts and U in each position is made by the fit of the line ratios.
 \Ts affects strongly the [NeIII] line, while [FeIII]/[FeII] is more
affected by U.

The  velocity V and  the density n are linked by the continuity
equation : V0 n0 = V1 n1.
Moreover, \B0, \n0,  and \Vs, are combined in the compression equation (Cox, 1972).
In conclusion,  all the parameters
are combined together in the calculations. Therefore, for each position a
large grid of models are run. The models are selected  on the
basis of the minimum deviation from the observation data for all
the line ratios.
\section{Spitzer observations}

\subsection{The line spectra}

\begin{figure}
\includegraphics[width=0.41\textwidth]{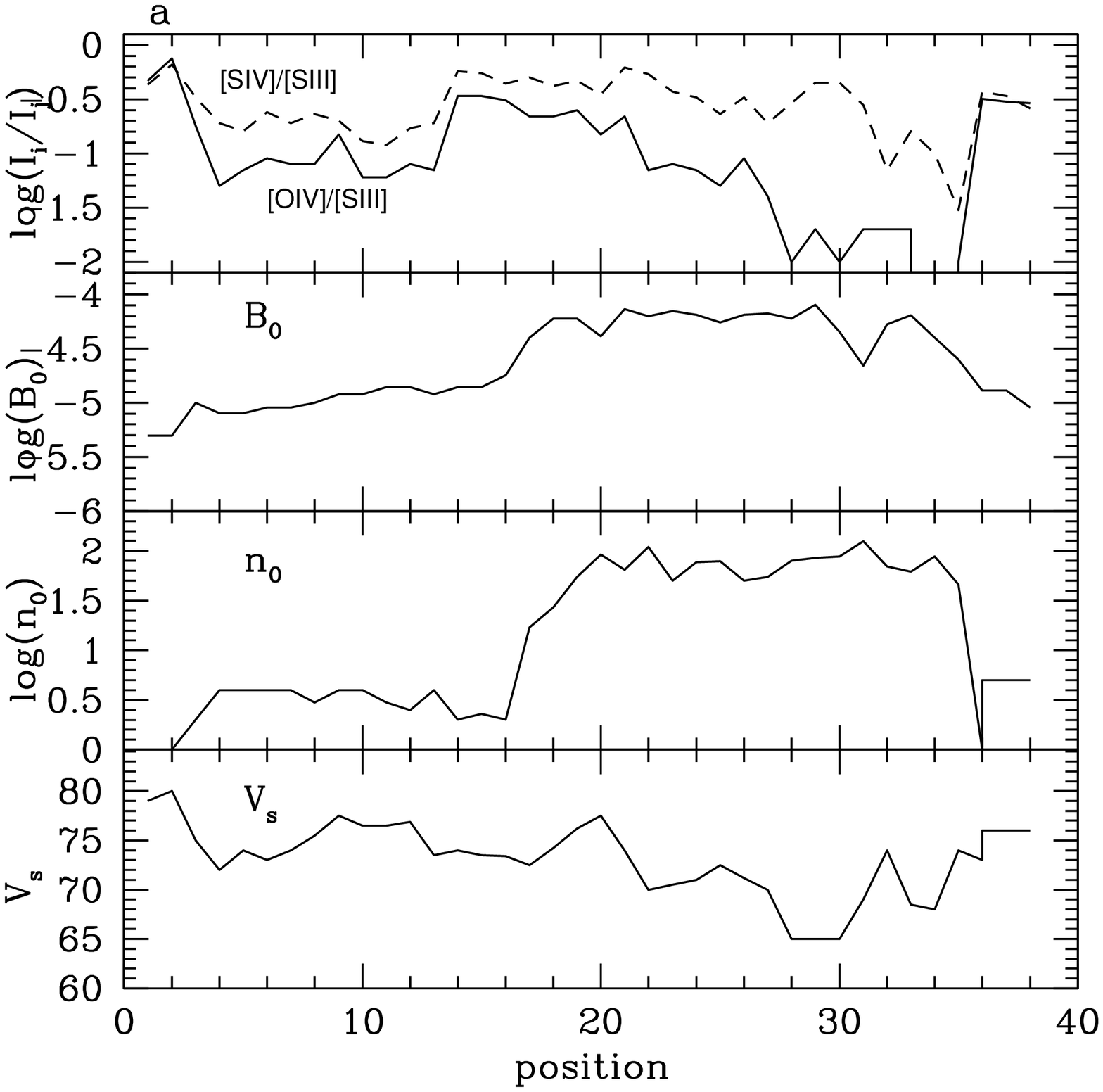}
\includegraphics[width=0.41\textwidth]{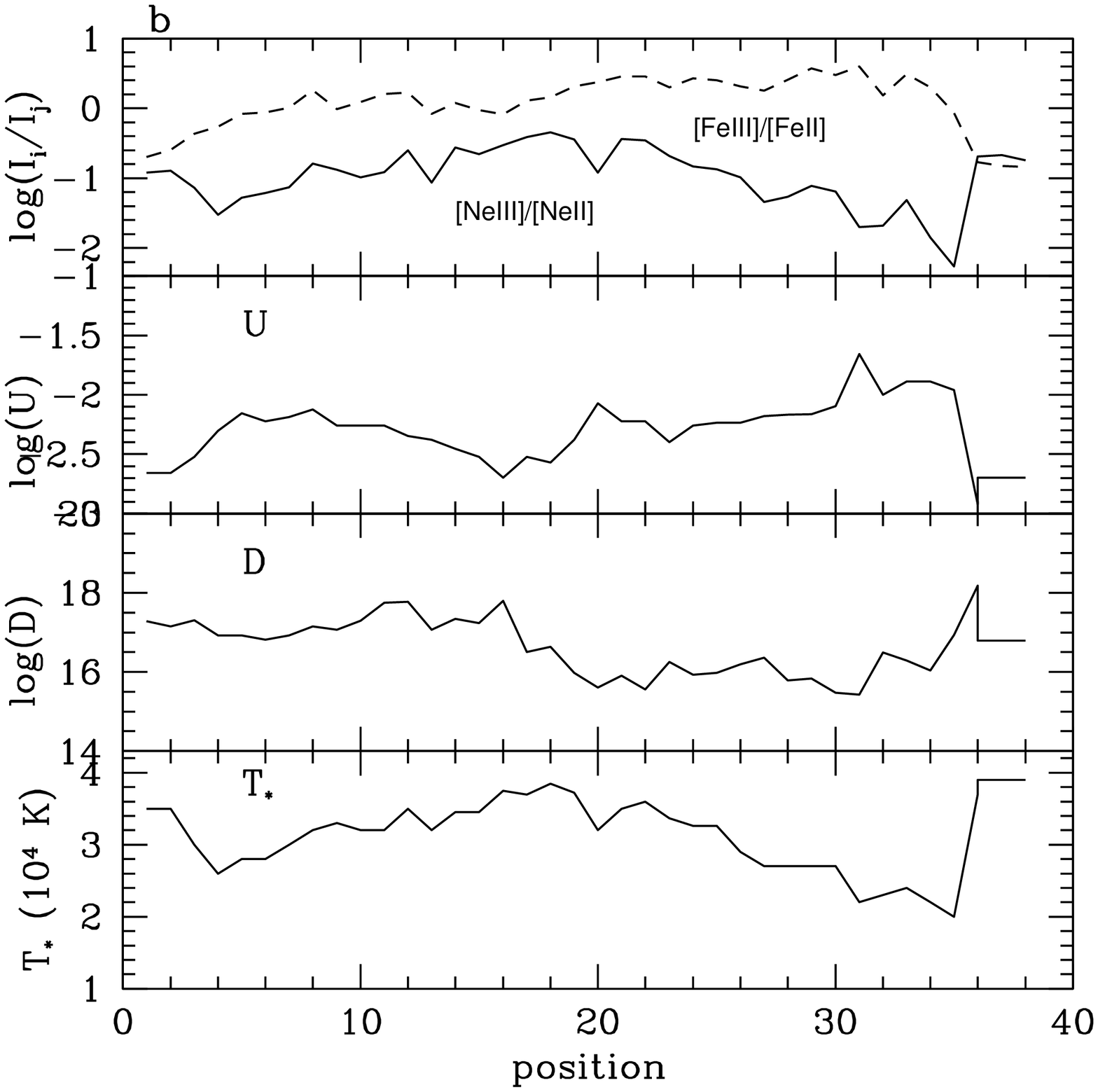}
\includegraphics[width=0.41\textwidth]{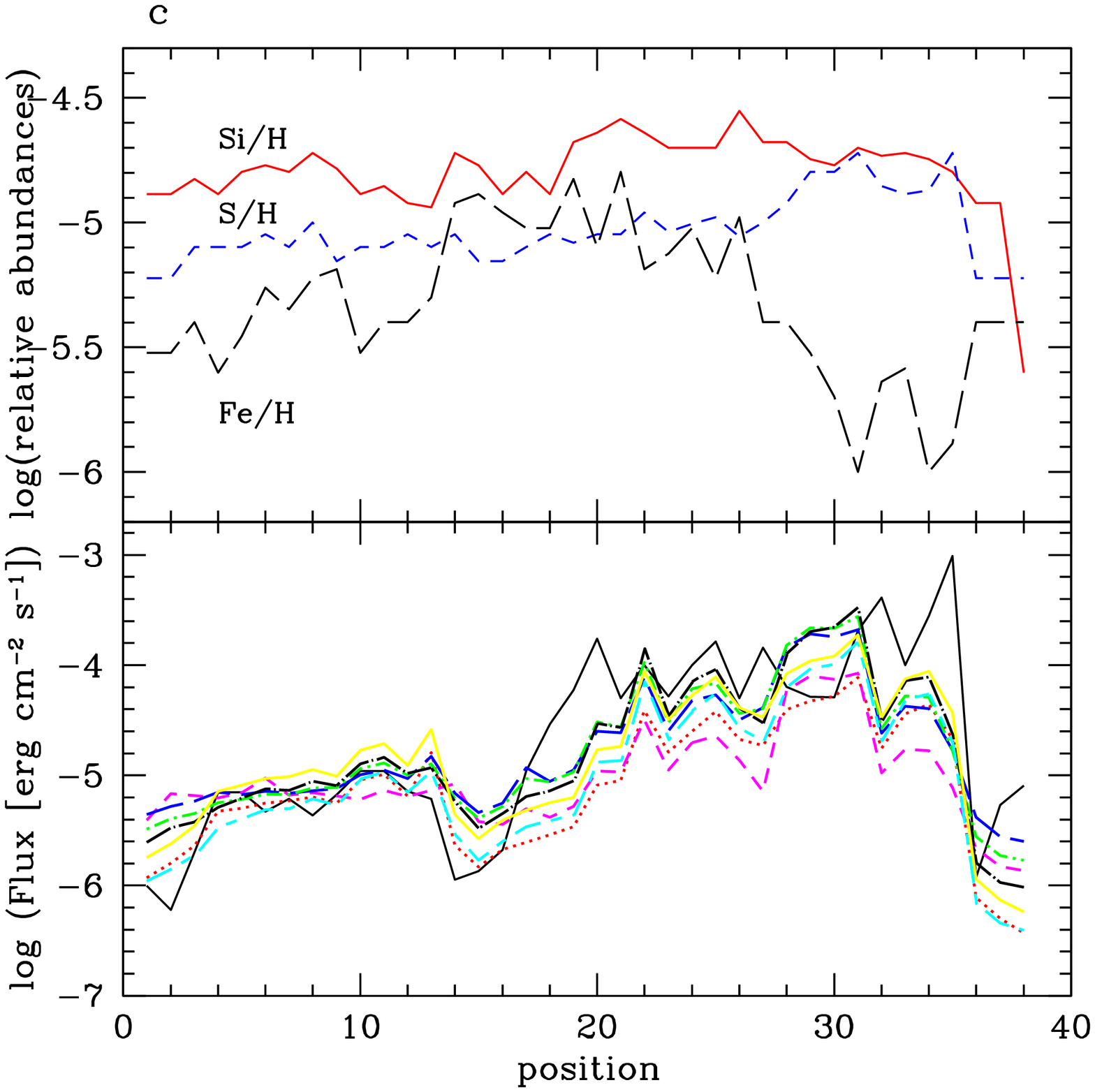}
\caption{The profiles of the different parameters which result from  modelling. a :the parameters depending on the shock;
 b : the parameters depending on photoionization;
c : the relative abundances (top panel); comparison of the continuum fluxes at different wavelength
with the calculated bremsstrahlung in the different positions (bottom panel);
magenta : 10.0-10.48 \mum, blue : 13.5-14.3 \mum, green : 15.56 \mum, black : 18.71 \mum, cyan : 22.29 \mum,, yellow : 28.22 \mum,
red : 33.48 \mum
 Each parameter units appear in Table 2}.
\end{figure}

 Spitzer  MIR spectra  (Program 3295, all AORs) were observed
 by  Simpson et al. (2007, hereafter S07) in   
 38 positions along a line approximately perpendicular to the Galactic plane in the GC (Fig. 1).
The line intensities were corrected by the $\tau_{\lambda}$/$\tau_{9.6}$  optical depth ratios,
using  the optical depth at 9.6 \mum $\tau_{9.6}$, as given by S07  in their Table 2 and 
Table 1 respectively. 

In some positions, in particular between 1 and 16, the observed [SIII]18.7/[SIII]33.5 ratios,
corrected for extinction by the $\tau_{9.6}$  presented in S07, were 
lower than the  calculated [SIII]18.7/[SIII]33.5  lower limit ($\sim$ 0.5),
even adopting a very low density model.
Consequently,   
the $\tau_{9.6}$ values were  incremented in each position in order to lead to a  reasonable spectrum.
.  In fact, the $\tau_{9.6}$ values were calculated
by Simpson et al (2007) assuming \Te=6000 K, while in our models the temperatures
downstream, depending on the shock velocity, show higher values in the region close to the shock front
(Sect. 3.2.1).

In Table 1, we compare  the spectra corrected for extinction with model calculations. 
In the last column the  extinction is given.
The  spectra are  numerated   from 1 to 38  referring to S07. 
Each observed (corrected) spectrum is followed in
the next  row by the calculated one, which  is numerated  from m1 to m38.
 Model m$_{pl}$ which appears in the last row of Tables 1 and 2 is explained in Sect. 3.3.

The models  adopt a black body photo-ionizing radiation flux  corresponding to the
colour temperature of the stars.  The model parameters are given in Table 2, where 
columns 2-4  list the shock parameters. Columns 5 and 6  give the photoionizing flux : the temperature of
the stars and the ionization parameter, respectively. The relative abundances 
Si/H, S/H, and Fe/H follow in columns 7-9. The O/H and Ne/H  ratios were found 
nearly constant in all S07 positions, by a first modelling. 
 In fact, depletion into dust grains is not important since 
O is not the main  constituent of dust grains and Ne cannot be adsorbed
due to its atomic structure.
The last column shows the geometrical thickness
of the emitting filaments. Indeed,  a high fragmentation of matter appears in the observed region
and in each position,  many different conditions could coexist. In our modelling
we refer to the data as to  single (average)  spectra.

In Table 1, we    show the line ratios normalized to [SIII] 33.5 =10 -the strongest line -
to avoid very small values.
The line sequence  is  ordered by  wavelength.
A grid of models was run  for each position and  the best fitting spectrum  was selected  
on the basis   of  the [SIV]/[SIII], [NeIII]/[NeII], and [FeIII]/[FeII] flux ratios
which do not depend on the relative abundances, and  of [OIV]/[SIII] which depends strongly on the
shock velocity.

To understand the results, we recall that  the radiative ionization rates  depend on
the intensity of the primary and secondary (diffuse) radiation  flux; 
  radiation cannot heat the gas to  $>$ 2-3 10$^4$ K.  
The shocks heat the gas to temperatures which depend on the
shock velocity and the collisional ionization rates 
increase with increasing temperatures. 

We   derive  \Ts and U  by the best fit of [NeIII]/[NeII] and  [FeIII]/[FeII], respectively.
However, these parameters   also affect [SIV]/[SIII].  So the whole process is   iterated until
all the line ratios are reproduced.

The  [OIV] line   corresponds to a relatively high ionization level
that  can be reached collisionally by a  relatively high temperature gas,
depending on \Vs.  
It was found that the [OIV] line
is  hardly detected  for shock velocities $<$ 64 \kms. 
Only shocks  with \Vs $>$ 64 \kms can  lead to results suitable to the observations. 

The ionization potential of  S$^{+3}$ is  lower than those
of  O$^{+2}$ and Ne$^{+2}$.  
Therefore,  the [SIV] line intensity   depends  on U and \Ts more than on \Vs.
Moreover,  [SIV]/[SIII]  decreases with distance from the shock front downstream 
following recombination, because  the S$^{+3}$ region
 is totally included within the nebula, while the S$^{+2}$ region can be cut off at a certain distance from
the shock front in matter-bound nebulae.
The geometrical thickness of the emitting cloud is  determined  when  the calculated   
[SIV]/[SIII] line ratio reproduces  the observed one.
The relative abundance of S is determined by  all the line ratios because they are given as ratios to
[SIII]. When  the line ratios  are  either all overestimated or all underestimated
by the same factor, S/H  is  modified in order to reproduce the data.  S  and
 Si  are  not strong coolant,  therefore  Si/H and S/H result directly, without re-starting the
whole  modelling process.

\subsection{Results}

The results of modelling are presented in the three diagrams of  Fig. 2. a,b,c.
We define as {\it results} the  sets of input parameters (\Vs, \n0,
\B0, \Ts, U, D, and the relative abundances) which lead to the best fit
of the  data in each position. 

 When a cloud moves toward the photoionization source, the shock front edge is reached directly
by the photons. When the cloud propagates outwards, the photoionising flux reaches the cloud
on the edge opposite to the shock front. Therefore, the calculations need some iterations.
In the present modelling, the best fit is obtained  considering that the shock front is reached by radiation 
from the hot stars. This
indicates that the clouds move towards the photoionization source.
The case of an outward motion  was discarded because we could not reach
a consistent fit of all the lines in  the different positions.

Comparing our results with those generally derived from
specific line ratios (e.g. Simpson et al. 2007), we recall
 that  \n0 and \B0 are  pre-shock values, while electron densities and temperatures
  derived from the  observed line ratios, refer to the values in the downstream regions.

To illustrate the results, we  present  in Fig. 3 the  profiles of
the electron temperature, the electron density, and
 the fractional abundance of the most significant ions downstream, e.g. for model m18.

\subsubsection{The parameters depending on the shock}

\begin{figure}
\begin{center}
\includegraphics[width=0.45\textwidth]{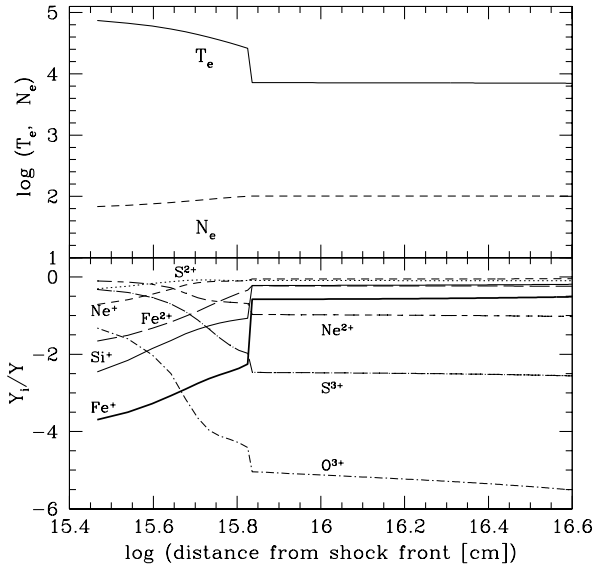}     
\caption{The distribution  (top panel) of
the electron temperature, the electron density, and (bottom panel)
of  the fractional abundance of the most significant ions downstream for model m18.
}
\end{center}
\end{figure}

In  Fig. 2a, \Vs, \n0, and \B0 are  shown as a function of  position.
The curves are not smooth, because the matter is strongly fragmented
and the calculations refer to {\it averaged} observations in each position.
There is a trend of decreasing  \Vs from positions 1 to 30,  with a   fluctuating increase  
from  30   to 38.
The  density is minimum between 14 and 16, namely in the Bubble. The  abrupt increase  in \n0 by a 
factor  $\leq$ 10 after position 16, leads to relatively high densities  up to position 35,
which corresponds to the limit of the Arched Filament region. Then the density returns to 
the values characteristic of the ISM.

The trend of  increasing  \B0  starts smoothly from position 1 and follows, on a reduced scale, the bump of
the density in the Arched Filament region. 
Considering that line excitation and recombination coefficients,  and cooling and heating
rates depend on the density of the gas downstream,  \B0  and \n0 must be cross-checked in each position. 

The magnetic field throughout the regions covered by the slit (Fig. 1) is still under discussion (S07).
Yusef-Zadeh \& Morris (1987a) claim
that the magnetic field in the linear non-thermal filaments of the Radio Arc is as high as  10$^{-3}$ gauss, 
while  LaRosa et al (2005) found that the global field of the GC is only $\sim$ 10$^{-5}$
gauss with an upper limit of $\leq$ 10$^{-4}$ gauss. 
The low value is similar to that in the ISM. The consistent modelling of all the lines
in each position leads to a growth of \B0 in the arched filament region by a factor of 10
(from $\sim$ 10$^{-5}$ to $\sim$ 10$^{-4}$ gauss), while in the surrounding ISM, \B0 is lower.

In  Fig. 2a (top panel), we notice that  the observed [OIV]/[SIII] line ratio
 follows the  \Vs decreasing slope, while [SIV]/[SIII]
is affected also by   the other parameters. 

\subsubsection{The parameters depending on the radiation flux}

In the bottom panel of  Fig. 2b, the temperature of the stars,
which are the source of the photoionizing flux,  is shown as a function of position.
A  maximum corresponding to $\sim$ 3.7 10$^4$ K appears in the region corresponding to
the Bubble. These  temperatures  refer to the Quintuplet cluster.
Star  temperatures  in the Arches Cluster are lower, $\leq$ 3 10$^4$ K.
We would expect a higher U in the positions  closest to the clusters, but 1)  we are dealing
with projected distances
and 2) the geometrical thickness   has a crucial role in reducing the photoionization flux
throughout the different slabs downstream. 
Fig. 2b in fact shows that the  minimum of U  between
positions  14 and 20 is accompanied by a rough bump in D.

Fig. 2b shows that
 the observed [NeIII]/[NeII] line ratio is correlated with \Ts,
while [FeIII]/[FeII] is roughly proportional to U.

\subsubsection{The relative abundances}

The relative abundances resulting from modelling
are shown  in  Fig. 2c (top panel). They are calculated consistently with
\Vs, \n0, \B0, and \Te

We had some difficulty to determine  O/H because the [OIV]/[SIII] line ratio could depend
 on \Vs ,  O/H, and on S/H  as well. We decided to keep O/H constant
 (and Ne/H constant) only after  considering a first iteration of results from all the positions
and  noticing  that O/H and Ne/H  were changing very little from position to position.
   The best results were obtained by O/H = 6. 10$^{-4}$
and Ne/H = 10$^{-4}$ (Ne/O=0.167) and  N/O= 0.18.  Grevesse \& Sauval (1998) find  Ne/O=0.15 and
N/O=0.123. 
These values are within the range calculated by Mart\'{i}n-Hern\'{a}ndez et al (2002)
for HII regions in the Galaxy.
 For elements  whose lines
were not observed,  we  adopted   the following  abundances : C/H=3.3 10$^{-4}$, N/H= 1.1 10$^{-4}$,
Mg/H = 2.6 10$^{-5}$, and Ar/H = 6.3 10$^{-6}$ (Allen 1973).
Thus, they do  not appear in Table 2.

 Si appears through one only line. 
Si/H  is determined  in each position after  the satisfactory fit of all the other line ratios.
In fact Si is not a strong coolant.
 The Si/H  relative abundance  can  assume all the
values below the solar one (3.3 10$^{-5}$), because Si is the main component of silicate grains.

The most impressive result refers to Fe. The strong depletion of Fe from the gaseous phase in the
Arched Filament  region indicates that iron is quite likely   depleted into  grains.
However,  its relative abundance is constrained
by  both [FeII] and [FeIII] lines, and therefore cannot  always be  derived directly by changing  Fe/H.  

 Small grains are easily sputtered. 
Iron returns to the gaseous  phase at higher \Vs  close to the Quintuplet Cluster.  
 Si  is  slightly depleted from the gaseous phase along all the slit positions   with some fluctuations,
indicating a minimum of silicate grains near the Sickle. 
Si/H  reaches values close to  solar in the Arched Filaments beyond  the Arches Cluster.
Perhaps  silicate grains are  evaporated by the strong radiation from the cluster.
In the Bubble, both Si/H and Fe/H are  slightly reduced, indicating that a large fraction of  grains survive. 
  Although the IR range observed  by Simpson et al.  includes the  PAH bands observed in the Milky Way (Draine \& Li 2007),
silicates including iron  and other iron species,
can also  be  present in dust. 
We will discuss these features   through  the analysis of the continua at different
wavelengths which  is shown  in the   bottom panel of  Fig. 2c,
after modelling the continuum SED in the next section.

\subsection{A power-law flux from the Galactic center?}

 The  results in the extreme positions 1-2, 36-38, showing  a relatively high \Ts, are
unusual  in the ISM,
suggesting that perhaps we  should try models which account for a power-law (pl) flux
from the center of the Galaxy, as for AGNs (e.g. Contini \& Viegas 2001), instead of a black body (bb). 
In fact, in the Galaxy center there is a "quiescent" BH (Ghez et al. 2005), 
which is  likely not far from the observed positions. 
We  have run a grid of models (m$_{pl}$) with a  pl ionization flux and the other parameters
similar to those found by modelling with a bb. The selected model appears in the last row
of Tables 1 and 2.
Actually, we could not find a  fit to the observed line ratios as good as that found  by the bb models.
In addition, a small contribution of this spectrum would not change the results.

The flux F$_{\nu}$ is
lower by at least a factor of 100 than the lowest flux found for LINERS (Contini 1997)
and for low  luminosity AGN (Contini 2004). 
Moreover, we found
O/H=8 10$^{-4}$ and Ne/H= 10$^{-4}$.

The  best fitting pl model was run with photoionization and shocks acting 
on the same edge of
the cloud, which means that the cloud is propagating towards the (supposed active) center. Inflow
is  characteristic of regions close to the  AGN  active center.

\begin{table*}
\begin{center}
\caption{Comparison with models of the observed line intensity ratios ([SIII] 33.5 =10) corrected for extinction}
\begin{tabular}{ l l l l l l l l l l l l} \\ \hline \hline
\  Position   & [SIV]10.5& [NeII]12.8& [NeIII]15.6&[SIII]18.7&[FeIII]22.9&[OIV]25.9&[FeII]26 & [SIII]33.5 & [SiII]34.8 &ext(9.6\mum)\\ \hline
\    1&  0.43& 14.94&  1.80&  5.15&  0.27&  0.47&  1.33& 10.00& 28.56 & 1.44\\
\  m1 & 0.43 &14.8  & 1.7  &5.29  &  0.25&  0.48&  1.2 &10.   &28.56 & \\
\    2&  0.66& 13.50&  1.74&  5.20&  0.41&  0.75&  1.61& 10.00& 27.57& 2.29\\
\  m2 & 0.7  &13.3  & 1.74 &  5.29& 0.34 &  0.74&  1.77& 10.  & 27.4 &\\ 
\    3&  0.33& 10.63&  0.78&  5.30&  0.38&  0.18&  0.88& 10.00& 19.61& 2.18\\
\  m3 & 0.33 &10.1  & 0.82 &  5.3 & 0.37 &  0.18&  0.83& 10.  & 19.8& \\ 
\    4&  0.19& 12.36&  0.37&  5.44&  0.31&  0.05&  0.57& 10.00& 18.53& 2.09\\
\  m4 &  0.184&12.0 & 0.39 &  5.47& 0.30 & 0.058&  0.59& 10.  & 18.3&  \\ 
\    5&  0.16&  9.04&  0.48&  5.48&  0.45&  0.07&  0.54& 10.00& 14.64& 2.37\\
\ m5  & 0.16 & 9.48 & 0.46 &  5.5 & 0.45 & 0.07 &  0.44& 10.  & 14.3& \\  
\    6&  0.24&  8.57&  0.53&  5.50&  0.60&  0.09&  0.69& 10.00& 14.97&2.83\\
\ m6  & 0.25 & 8.5  & 0.53 &  5.47& 0.6  & 0.08 &  0.67& 10.  & 14.4&\\
\    7&  0.19&  8.43&  0.63&  5.39&  0.50&  0.08&  0.49& 10.00& 12.61&2.34\\
\ m7  & 0.19 &  8.7 & 0.67 &  5.48& 0.55 & 0.08 &  0.47& 10.  & 12.9& \\
    8&  0.23&  5.78&  0.94&  5.46&  0.58&  0.08&  0.32& 10.00&  8.78& 2.58\\
\ m8  & 0.23 &  6.0 & 0.93 &  5.4 &  0.56& 0.08 &  0.25& 10.  & 8.4& \\  
\    9&  0.20&  9.26&  1.23&  5.46&  0.83&  0.15&  0.85& 10.00& 14.59& 2.32\\
\ m9  & 0.21 & 9.29 & 1.28 &  5.46&  0.85& 0.146&  0.7 & 10.  & 14.7&\\
\   10&  0.13&  8.20&  0.84&  5.53&  0.38&  0.06&  0.31& 10.00& 10.50&2.30\\
\ m10 & 0.13 & 8.38 & 0.84 &  5.45& 0.35 & 0.065& 0.3  & 10.  & 10.6& \\
\   11&  0.12&  7.44&  0.91&  5.41&  0.42&  0.06&  0.26& 10.00&  9.28&2.46\\
\ m11 & 0.12 & 7.8  & 0.9  &  5.38& 0.477& 0.06 &  0.278&10.  & 9.24& \\
\   12&  0.17&  6.09&  1.53&  5.31&  0.42&  0.08&  0.25& 10.00&  7.17&2.32\\
\ m12 & 0.17 & 6.6  & 1.44 &  5.35&  0.4 & 0.08 &  0.25& 10.  &  7.6&\\
\   13&  0.19&  9.51&  0.82&  5.44&  0.52&  0.07&  0.62& 10.00& 10.36& 2.54\\
\ m13 & 0.19 & 9.   & 0.84 &  5.44& 0.54 & 0.076& 0.68 & 10.  & 11.3& \\
\   14&  0.57&  6.28&  1.74&  5.22&  1.06&  0.34&  0.88& 10.00& 12.54&2.75\\
\ m14 & 0.57 & 6.47 & 1.79 & 5.3  & 1.1  & 0.32 & 0.83 & 10.  & 12.5& \\
\   15&  0.55&  9.46&  2.10&  5.21&  1.42&  0.34&  1.51& 10.00& 16.91&1.75\\
\ m15 & 0.56 & 9.1  & 2.1  &  5.3 & 1.47 & 0.34 & 1.6  & 10.  & 17.2& \\
\   16&  0.44&  8.51&  2.54&  5.24&  1.12&  0.31&  1.37& 10.00& 14.10&1.57\\
\ m16 & 0.44 & 8.7  & 2.65 &  5.29& 1.19 & 0.34 & 1.43 & 10.  & 14.1& \\
\   17&  0.55&  7.24&  2.81&  5.59&  0.95&  0.22&  0.73& 10.00& 11.75&1.75\\
\ m17 & 0.6  & 7.6  & 2.83 &  5.8 & 1.   & 0.23 & 0.77 & 10.  & 11.2& \\
\   18&  0.42&  6.49&  2.95&  6.05&  0.96&  0.22&  0.66& 10.00&  8.96&1.74\\
\ m18 & 0.45 & 6.8  & 2.86 &  6.  & 0.93 & 0.22 & 0.69 & 10.  & 8.04& \\
\   19&  0.47&  8.74&  3.12&  7.34&  1.91&  0.25&  0.93& 10.00& 11.45&1.80\\
\ m19 & 0.47 & 8.6  & 3.16 &  7.32& 1.9  & 0.24 & 1.16 & 10.  & 11.48&\\
\   20&  0.35& 12.04&  1.44& 10.75&  1.45&  0.15&  0.61& 10.00&  8.96&2.59\\
\ m20 & 0.34 & 12.1 & 1.5  & 10.7 & 1.47 & 0.157& 0.69 & 10.  & 8.5& \\
\   21&  0.62&  7.50&  2.75&  7.31&  1.67&  0.22&  0.58& 10.00&  8.89&2.44\\
\ m21 & 0.62 & 7.5  &  2.8 & 7.3  & 1.7  & 0.22 & 0.55 & 10.  & 8.89& \\
\   22&  0.54&  8.00&  2.77&  9.72&  0.75&  0.07&  0.26& 10.00&  6.68&3.17\\
\ m22 & 0.54 & 8.   & 2.77 & 9.6  & 0.76 & 0.075 & 0.3 & 10.  & 6.63& \\
\   23&  0.37&  7.42&  1.55&  6.57&  0.86&  0.08&  0.43& 10.00&  9.53&2.50\\
\ m23 & 0.37 & 7.5  & 1.6  & 6.66 & 0.83 & 0.08 & 0.43 & 10.  & 9.3&\\
\   24&  0.33&  8.18&  1.20&  8.38&  1.22&  0.07&  0.45& 10.00&  7.83&3.11\\
\ m24 & 0.31 & 8.3  & 1.34 & 8.1  & 1.2  & 0.07 & 0.57 & 10.  & 7.6&\\
\   25&  0.23&  8.54&  1.15&  8.73&  0.78&  0.05&  0.31& 10.00&  7.50&2.91\\
\ m25 & 0.23 & 8.6  & 1.16 & 8.62 & 0.75 & 0.05 & 0.42 & 10.  & 7.7& \\
\   26&  0.33&  8.32&  0.86&  6.73&  1.31&  0.09&  0.63& 10.00& 11.93&2.67\\
\ m26 & 0.33 & 8.34 & 0.88 & 6.74 & 1.35 & 0.09 & 0.6  & 10.  & 11.84& \\
\   27&  0.19&  8.10&  0.37&  8.07&  0.50&  0.04&  0.28& 10.00&  7.52&2.91\\
\ m27 & 0.19 & 7.8  & 0.45 & 7.0  & 0.50 &  0.04 & 0.22& 10.  &7.7&   \\
\   28&  0.29&  7.22&  0.39&  7.95&  0.44&  0.01&  0.17& 10.00&  6.33&3.18\\
\ m28 & 0.29 & 7.3  & 0.45 & 7.9  & 0.44 & 0.014& 0.2  & 10.  & 6.3& \\
\   29&  0.45&  5.27&  0.41&  7.89&  0.26&  0.02&  0.07& 10.00&  3.39&3.11\\
\ m29 & 0.42 & 5.   & 0.46 & 7.5  & 0.23 & 0.02 & 0.09 & 10.  & 3.47& \\
\   30&  0.45&  6.51&  0.42&  8.78&  0.24&  0.01&  0.08& 10.00&  3.95&3.12\\
\ m30 & 0.4  & 6.2  & 0.42 & 8.9  & 0.18 & 0.01 & 0.1  & 10.  & 3.9 \\ \hline
\end{tabular}
\end{center}
\end{table*}

\begin{table*}
\begin{center}
\centerline{Table 1 (cont)}
\begin{tabular}{ l l l l l l l l l l l l} \\ \hline
\  Position   & [SIV]10.5& [NeII]12.8& [NeIII]15.6&[SIII]18.7&[FeIII]22.9&[OIV]25.9&[FeII]26 & [SIII]33.5 & [SiII]34.8 &ext.(9.6 \mum)\\
\   31&  0.28&  8.48&  0.17& 13.05&  0.16&  0.02&  0.04& 10.00&  2.92&3.40\\
\ m31 & 0.28 & 8.1  & 0.2  & 12.9 & 0.13 & 0.02 & 0.05 & 10.  & 2.7& \\
\   32&  0.07&  7.68&  0.16&  7.98&  0.26&  0.02&  0.17& 10.00&  6.28&2.51\\
\ m32 & 0.07 & 7.76 & 0.15 & 8.2  & 0.27 & 0.02 & 0.18 & 10.  & 6.& \\
\   33&  0.16&  5.37&  0.26&  7.12&  0.28&  0.02&  0.09& 10.00&  3.85&3.21\\
\ m33 & 0.16 & 5.9  & 0.23 & 7.14 & 0.26 & 0.02 &  0.07& 10.  & 3.2&\\
\   34&  0.10&  9.18&  0.13&  9.10&  0.14&  0.00&  0.07& 10.00&  4.21&3.02\\
\ m34 & 0.10 & 9.0  & 0.14 & 9.36 & 0.14 & 0.01 & 0.078& 10.  & 4.8& \\
\   35&  0.03& 11.02&  0.06&  7.72&  0.13&  0.01&  0.15& 10.00&  6.17&2.00\\
\ m35 & 0.03 & 10.8 & 0.05 & 8.   & 0.13 & 0.01 &  0.28& 10.  & 7.2& \\
\   36&  0.37& 12.99&  2.67&  5.40&  0.24&  0.32&  1.41& 10.00& 23.54&0.80\\
\ m36 & 0.4  & 13.2 & 2.2  & 5.24 & 0.3  & 0.34 & 1.4  & 10.  & 24.& \\
\   37&  0.34& 13.41&  2.87&  5.48&  0.21&  0.30&  1.39& 10.00& 23.67&0.42\\
\ m37 & 0.36 & 13.  &  2.7 & 5.5  & 0.37 & 0.3  & 1.3  & 10.  & 21.3& \\
\   38&  0.26& 14.82&  2.70&  5.95&  0.187& 0.29&  1.23& 10.00& 23.27&0.33\\
\ m38 & 0.27 & 14.8 & 2.2  & 5.6  & 0.2 & 0.2   & 1.   & 10.  & 25 \\ \hline
\ m$_{pl}$ & 1.9&12.7&3.4 &  5.3  & 0.1 & 0.36  & 2.1  & 10   &30  \\ \hline\\
\end{tabular}
\end{center}
\end{table*}

\begin{table*}
\begin{center}
\caption{The models}
\begin{tabular}{ l l l l l l l l l l l l} \\ \hline \hline
\  model  & \Vs & \n0 & \B0  & \Ts     & U $^1$      & Si/H    & S/H    &  Fe/H  & D \\
\   &(\kms) & (\cm3) & (gauss) & (K)  & -  &  -      & -      &   -    & (cm)\\ \hline
\ m1&   79&   1&   5e-6&   3.5e4 &  2.2e-3 &  1.3e-5 &  6.e-6 &  3.e-6 &  1.92e17\\
\ m2&   80&   1&   5e-6&   3.4e4 &  2.1e-3 &  1.4e-5 &  7.e-6 &  5.e-6 &  1.40e17\\
\ m3&   75&   2&   1.e-5&  3.e4  &  3.e-3  &  1.5e-5 &  8.e-6 &  4.e-6 &  2.e17\\
\ m4&   72&   4&   8.e-6&  2.6e4 &  5.e-3  &  1.3e-5 &  8.e-6 &  2.5e-6&  8.5e16\\
\ m5&   74&   4&   8.e-6&  2.8e4 &  7.e-3  &  1.6e-5 &  8.e-6 &  3.5e-6&  8.3e16\\
\ m6 &  73&   4&   9.e-6&  2.8e4 &  6.e-3  &  1.7e-5  & 9.e-6 &  5.5e-6&  6.6e16\\
\ m7&   74&   4&   9.e-6&  3.e4  &  6.5e-3 &  1.6e-5 &  8.e-6 &  4.5e-6&  8.3e16\\
\ m8&   75.5& 3&   1.e-5&  3.2e4 &  7.5e-3 &  1.9e-5 &  1.e-5 &  6.e-6 &  1.43e17\\
\ m9&   77.5& 4&   1.2e-5& 3.3e4 &  5.5e-3 &  1.65e-5&  7.e-6 &  6.5e-6&  1.18e17\\
\ m10&  76.5& 4&   1.2e-5& 3.2e4 &  5.5e-3 &  1.3e-5 &  8.e-6 &  3.e-6 &  1.95e17\\
\ m11&  76.5& 3&   1.4e-5& 3.2e4 &  5.5e-3 &  1.4e-5 &  8.e-6 &  4.e-6 &  5.6e17\\
\ m12&  76.9& 2.5& 1.4e-5& 3.5e4 &  4.5e-3 &  1.2e-5 &  9.e-6 &  4.e-6 &  6.e17\\
\ m13&  73.5& 4  & 1.2e-5& 3.2e4 &  4.2e-3 &  1.15e-5&  8.e-6 &  5e-6  &  1.18e17\\
\ m14&  74  & 2  & 1.4e-5& 3.45e4&  3.5e-3 &  1.9e-5 &  9.e-6 &  1.2e-5&  2.2e17\\
\ m15&  73.5& 2.3& 1.4e-5& 3.45e4&  3.e-3  &  1.7e-5 &  7.e-6 &  1.3e-5&  1.7e17\\
\ m16&  73.4& 2  & 1.8e-5& 3.75e4&  2.e-3  &  1.3e-5 &  7.e-6 &  1.1e-5&  6.3e17\\
\ m17&  72.5& 17 & 4.e-5 & 3.7e4 &  3.e-3  &  1.6e-5 &  8.e-6 &  9.5e-6&  3.16e16\\
\ m18&  74.2& 27 & 6.e-5 & 3.85e4&  2.7e-3 &  1.3e-5 &  9.e-6 &  9.5e-6&  4.3e16\\
\ m19&  76.2& 55  &6.e-5 & 3.72e4&  4.2e-3 &  2.1e-5 &  8.3e-6&  1.5e-5&  9.5e15\\
\ m20&  77.5& 92 & 4.1e-5& 3.2e4 &  8.5e-3 &  2.3e-5 &  9.e-6 &  8.e-6 &  4.e15\\
\ m21&  74  & 65 & 7.3e-5& 3.5e4 &  6.e-3  &  2.6e-5 &  9.e-6 &  1.6e-5&  8.e15\\
\ m22&  70  & 110& 6.3e-5& 3.6e4 &  6.e-3  &  2.3e-5 &  1.1e-5&  6.5e-6&  3.6e15\\
\ m23&  70.5& 50 & 7.e-5 & 3.37e4&  4.e-3  &  2.e-5  &  9.2e-6&  7.5e-6&  1.8e16\\
\ m24&  71  & 77 & 6.5e-5& 3.26e4&  5.5e-3 &  2.e-5  &  9.9e-6&  9.5e-6&  8.6e15\\
\ m25&  72.5& 79 & 5.5e-5& 3.26e4&  5.8e-3 &  2.e-5  &  1.05e-5& 6.e-6 &  9.5e15\\
\ m26&  71.2& 50 & 6.5e-5& 2.9e4 &  5.8e-3 &  2.8e-5 &  8.8e-6 & 1.05e-5& 1.57e16\\
\ m27&  70  & 55 & 6.7e-5& 2.7e4 &  6.6e-3 &   2.1e-5&   1.0e-5&  4.e-6 & 2.3e16\\
\ m28&  65  & 80 & 6.e-5 & 2.7e4 &  6.8e-3 &  2.1e-5 &  1.2e-5 & 4.e-6  & 6.1e15\\
\ m29 & 65  & 85 & 8.e-5 & 2.7e4 &  6.9e-3 &  1.8e-5 &  1.6e-5 & 3.e-6  & 6.9e15\\
\ m30&  65  & 88 & 4.5e-5& 2.7e4 &  8.e-3  &  1.7e-5 &  1.6e-5 & 2.e-6  & 3.e15\\
\ m31&  69  & 125& 2.2e-5& 2.2e4 &  2.2e-2 &  2.e-5  &  1.9e-5 & 1.e-6  & 2.7e15\\
\ m32&  74  & 70 & 5.3e-5& 2.3e4 &  1.e-2  &  1.85e-5&  1.4e-5 & 2.3e-6 & 3.1e16\\
\ m33&  68.5& 62 & 6.4e-5& 2.4e4 &  1.3e-2 &  1.9e-5 &  1.3e-5 & 2.6e-6 & 1.95e16\\
\ m34&  68  & 88 & 4.e-5 & 2.2e4 &  1.3e-2 &  1.8e-5 &  1.35e-5& 1.e-6  & 1.1e16\\
\ m35&  74  & 46 & 2.5e-5& 2.e4  &  1.1e-2 &  1.6e-5 &  1.9e-5 & 1.3e-6 & 8.7e16\\
\ m36&  73  & 1  & 1.3e-5& 3.7e4 &  1.2e-3 &  1.2e-5 &  6.e-6  & 4.e-6  & 1.5e18\\
%\ m36&  76  & 5  & 1.3e-5& 3.9e4 &  2.e-3  &  1.2e-5 &  6.e-6  & 4.e-6  & 6.2e16\\
\ m37&  76  & 5  & 1.3e-5& 3.9e4 &  2.e-3  &  1.2e-5 &  6.e-6  & 4.e-6  & 6.2e16\\
\ m38&  76  & 5  & 9.e-6 & 3.9e4 &  2.e-3  &  2.5e-6 &  6.e-6  &2.5e-6  & 5.4e16\\ \hline
\ m$_{pl}$&75& 3 & 2.e-6 & $\alpha$=-2&F$_{\nu}$=1e6$^2$&8.e-6&  3.e-5  & 3.e-6& 4.2e16\\ \hline 
\end{tabular}
\end{center}

\flushleft

$^1$  U is a number

$^2$ in photons cm$^{-2}$ s$^{-1}$ eV$^{-1}$ at the Lyman limit.

\end{table*}

\subsection{The continuum}

\begin{figure}
\includegraphics[width=0.45\textwidth]{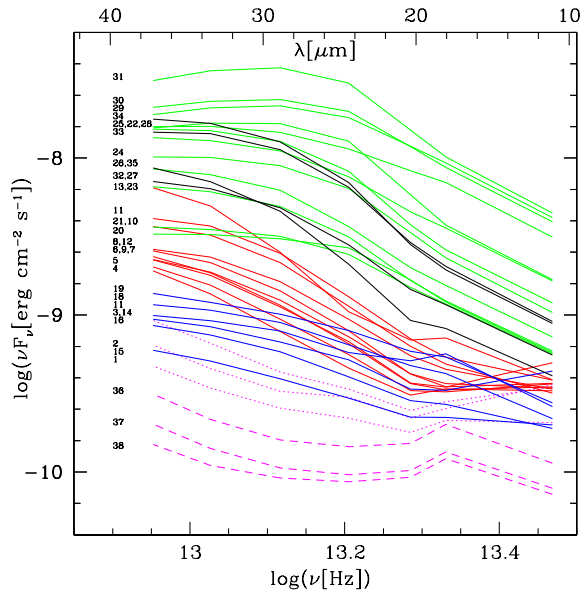}
\caption{The observed corrected continuum SEDs in the different positions.
magenta dotted : 1-3; red solid : 4-13; blue solid : 14-19;
green solid : 20-31; black solid : 32-35: magenta dashed : 36-38}
\end{figure}

\begin{figure*}
\includegraphics[width=0.3\textwidth]{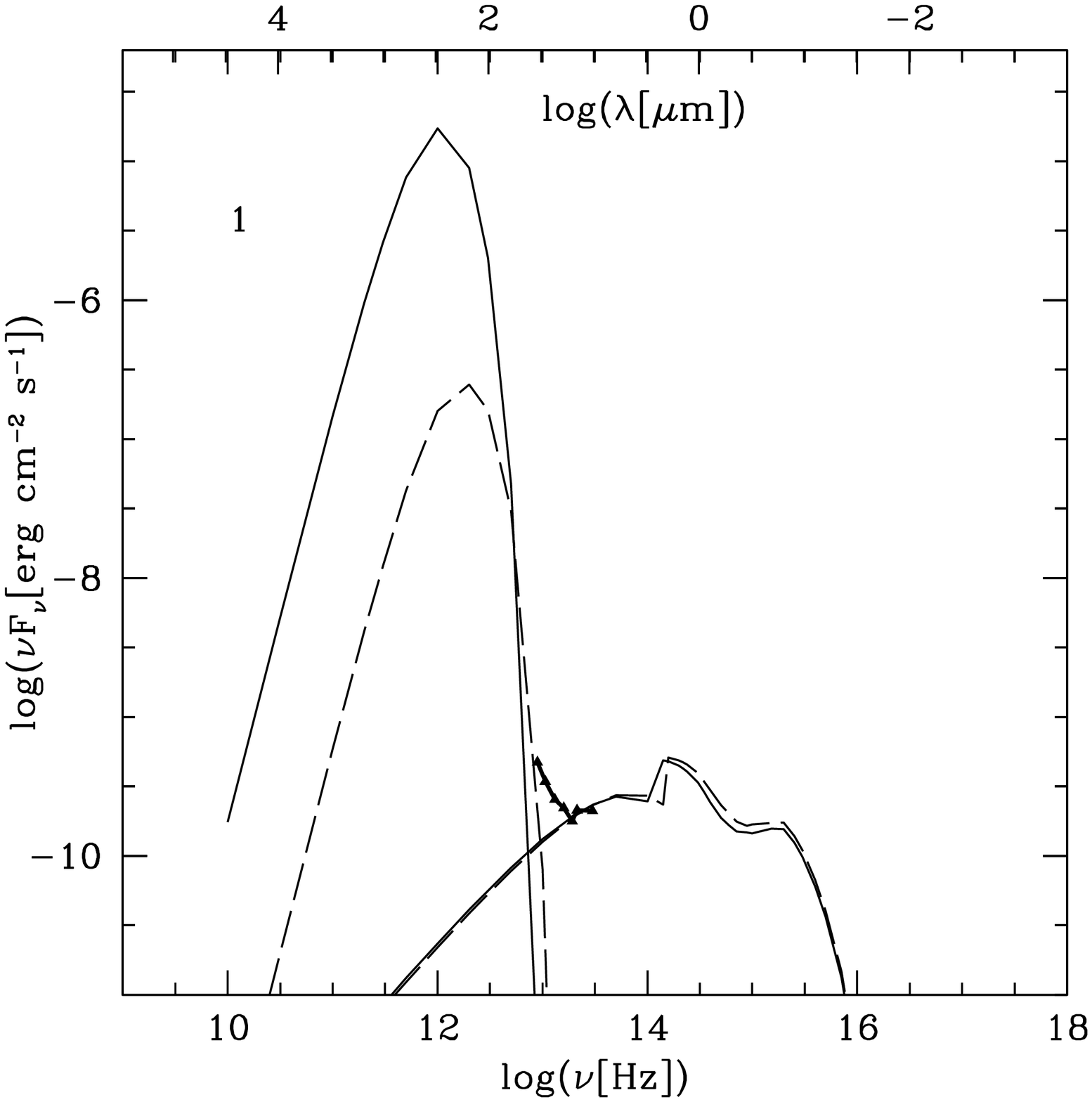}
\includegraphics[width=0.3\textwidth]{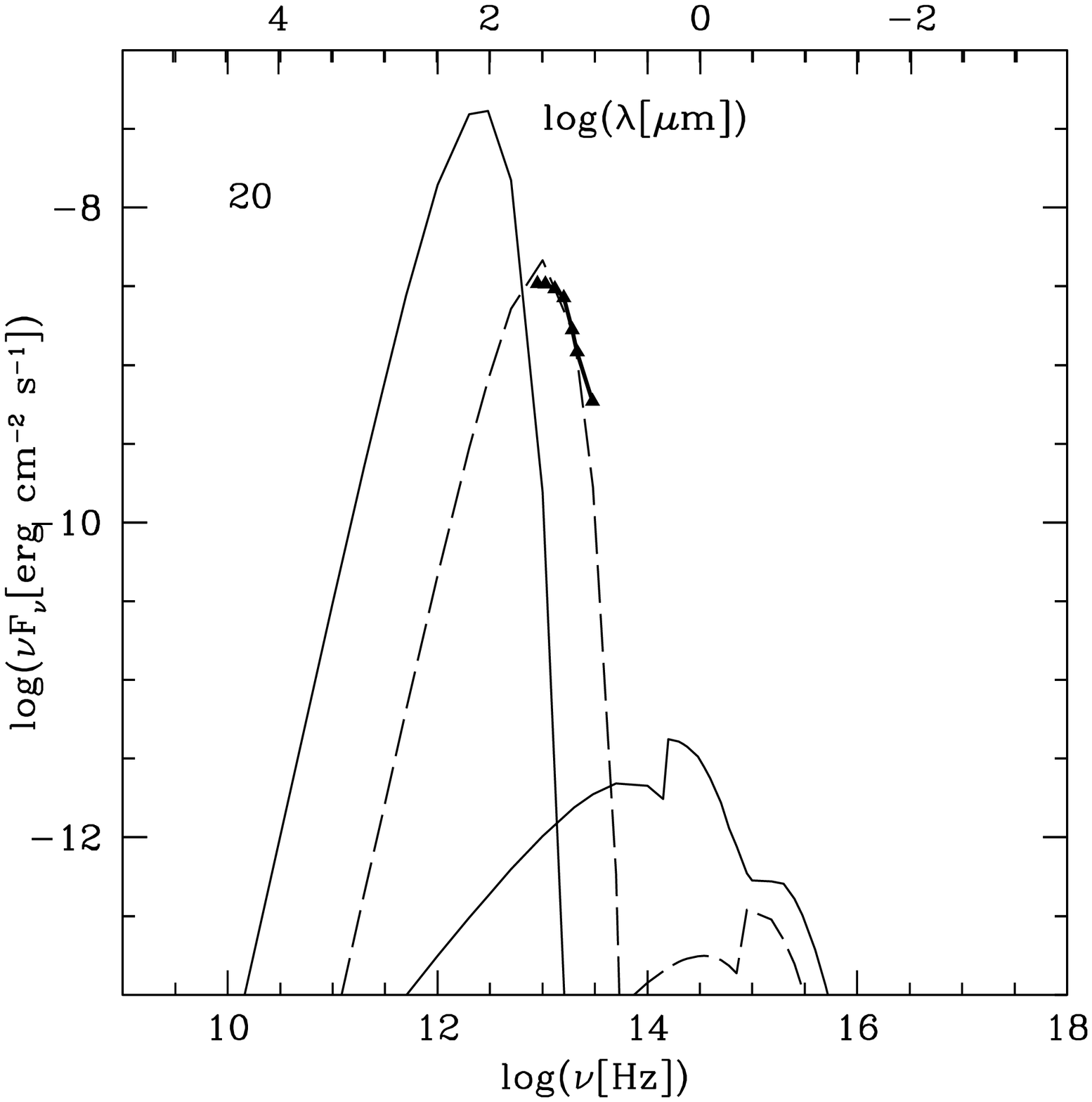}
\includegraphics[width=0.3\textwidth]{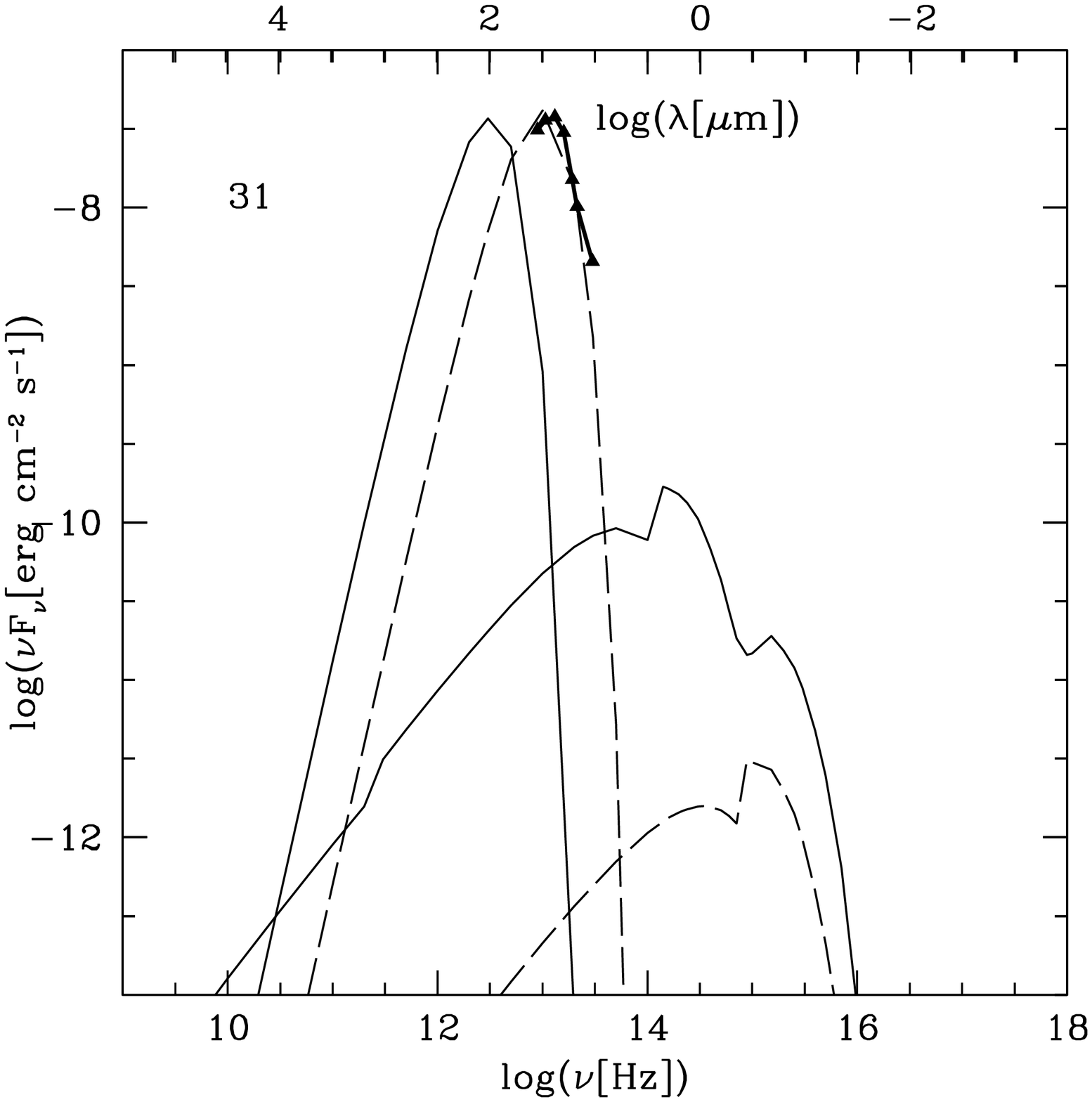}
\caption{Comparison of calculated with observed SEDs in positions 1, 20, and 31.
For each model two curves are shown : one referring to bremsstrahlung peaking at higher frequencies, 
and one peaking in the IR  referring to dust reprocessed
radiation. Solid  lines : models m1, m20, and m31 calculated by \agr=0.2 \mum. 
Dashed lines : in the left diagram correspond  to model m1 calculated by \agr=0.01
\mum, in the middle and right diagrams correspond to models calculated by
U=2.2. 
}
\end{figure*}

We  adopt the observational data  of the continuum by Simpson et al (2007, Table 4).
The data  were corrected for extinction  by the same correction parameters as those used for
the lines (Table 1). The SEDs in the different positions as a function of the frequency are shown
in Fig. 4.  The errorbands  are not included in the figure for the sake of clarity.
Two different  trends appear : one characteristic of   positions in the ISM : 1, 2, 3 (dotted lines) and 36, 37, 38
(dashed lines).
The other trend   prevails in the other regions (except in position 20). 
Different colours refer to different groups of positions,
in order to have a qualitative view of the SEDs.

\subsubsection{The continuum  SEDs}
To understand the trends in Fig. 4,  we present
in Fig. 5  the comparison of the continua calculated by the models which were used to calculate the line ratios
with the data for position 1, 20, and 31  in a  large  range of frequencies,  from radio to X-ray.   
Unfortunately, the data  cover only  the range  between $\sim$ 10 and 35 \mum.   
The calculated continuum SED shows the contribution of the gas (free-free+free-bound) and that of dust reprocessed
radiation, separately. 

An interesting  result appears from  Fig. 5, namely,  the 
  dust reradiation  peak predicted by the models, which  explain the line ratios (solid lines), 
occurs at a frequency 
lower than that  derived from  the  observations  in positions 20 and 31, while in position 1,  model m1 can 
 reproduce  the continuum  adopting grains with a radius  \agr $\sim$ 0.01 \mum (dashed lines).
 Very small grains
 can be heated stochastically to   temperatures  ($\leq$ 50 K, Draine 2003) which, however,
are not high  enough to shift the frequency peak.
PAHs correspond to small grains ($\leq$ 0.01 \mum), while the size of grains including Fe is still
under discussion (Hoefner 2009).

Previous calculations of models including dust (Contini et al 2004) show that the peak shifts to 
higher frequencies  1) increasing   \Vs  i.e. increasing the  collisional heating of dust grains,
2) increasing   U,  i.e. increasing radiation heating of the grains,  and 3)  reducing the radius of the grains.
Excluding collisional heating  derived from a higher velocity which  would imply   very  broad  line
profiles, the only parameters that we can alterate, are U and \agr. 
We have calculated some models with a very high ionization parameter which represent the case 
 of  an emitting cloud  closer to the  hot source, or less screened from the radiation flux.
For positions 20 and 31 we had to use a ionization parameter higher  by a factor $>$ 100  
than  that used for the lines  in order to fit the IR continuum data.
The  model, leading to the hot dust component,    produces  different line fluxes 
 destroying  the nice fit of the line ratios to the data shown in Table 1. So  this model contribution  corresponds
to  a low relative weight.

 In  positions 20 and 31,   a dust    
temperature of $\sim$ 150 K (dashed lines)  explains the data in the IR, 
while dust within the cloud emitting the 
line spectrum at position 20 reaches only a temperature of $\sim$ 38 K (solid lines).
Moreover, Fig. 5 shows that the IR continuum  in positions 20 and 31 is emitted  by dust
while in position 1 the data  are reproduced by the sum of reradiation fluxes by dust and bremsstrahlung. 
This explains the different slopes in Fig. 4. 
In agreement
with  very non-homogeneous matter in each observed region,  different
clouds contribute to the spectra. 
Alternatively,  
the relatively hot dust could be  spread in the central region of the Galaxy, independently from the gas morphology.

\subsubsection{Comparison of  IR continuum fluxes} 
In Fig. 2c (bottom  panel)  the  bremsstrahlung (black solid line) in the IR range {\it calculated
at the nebula} at each position,
is compared with the fluxes  corresponding to different wavelengths 
{\it observed (corrected) at Earth}, 
in the continuum.
 They are shifted  by a factor $\eta$  which  depends on the  distance of
the clouds from the photoionization source and  on the  distance of the clouds to Earth.
Adopting a distance to Earth of 8 kpc (Simpson et al 2007), the distance of the dusty clouds
from the cluster is  $\geq$ 30 pc.
Recall that both the  bremsstrahlung and the IR fluxes  depend on n$^2$ D (where n is the density
of the emitting gas), while the IR fluxes  between 10 and 33.48 \mum depend also on
the gas-to-dust ratios, because they  are generally emitted from reprocessed dust.
 A  perfect fit of  the bremsstrahlung with the IR observed fluxes
is not  expected due to the approximations of modelling. 
%Nevertheless, some major discrepancies as, e.g. in
%positions 19-20, 27-34  can be explained  by  the different $d/g$.  In other words, the maxima and minima of
%Fe/H in the top panel of Fig. 3c, are opposite to those of the IR fluxes in the bottom panel. 
%In fact, Fe/H  results  from the analysis of the  lines and  therefore refers to the
%gaseous phase, while the IR fluxes  refer to the continuum  emitted by dust.

Fig. 2c  shows that the  bremsstrahlung and the IR fluxes   have roughly similar profiles,
except  in the ISM  positions : in the  southern  positions, dust reradiation is higher than the bremsstrahlung
confirming that that Si, S, and Fe  could be depleted from the gaseous phase into grains,
while in the northern positions, the dust-to-gas ratios are low. 
%The only element which recovers
%from depletion in the gaseous phase is iron, while Si and S have an opposite trend.
%This would  confirm that silicates  and molecules including S, are present in the northern  ISM region.

\section{Position C - G0.095+0.012 and the E2 Thermal Radio Filaments}

\begin{figure}
\begin{center}
\includegraphics[width=0.45\textwidth]{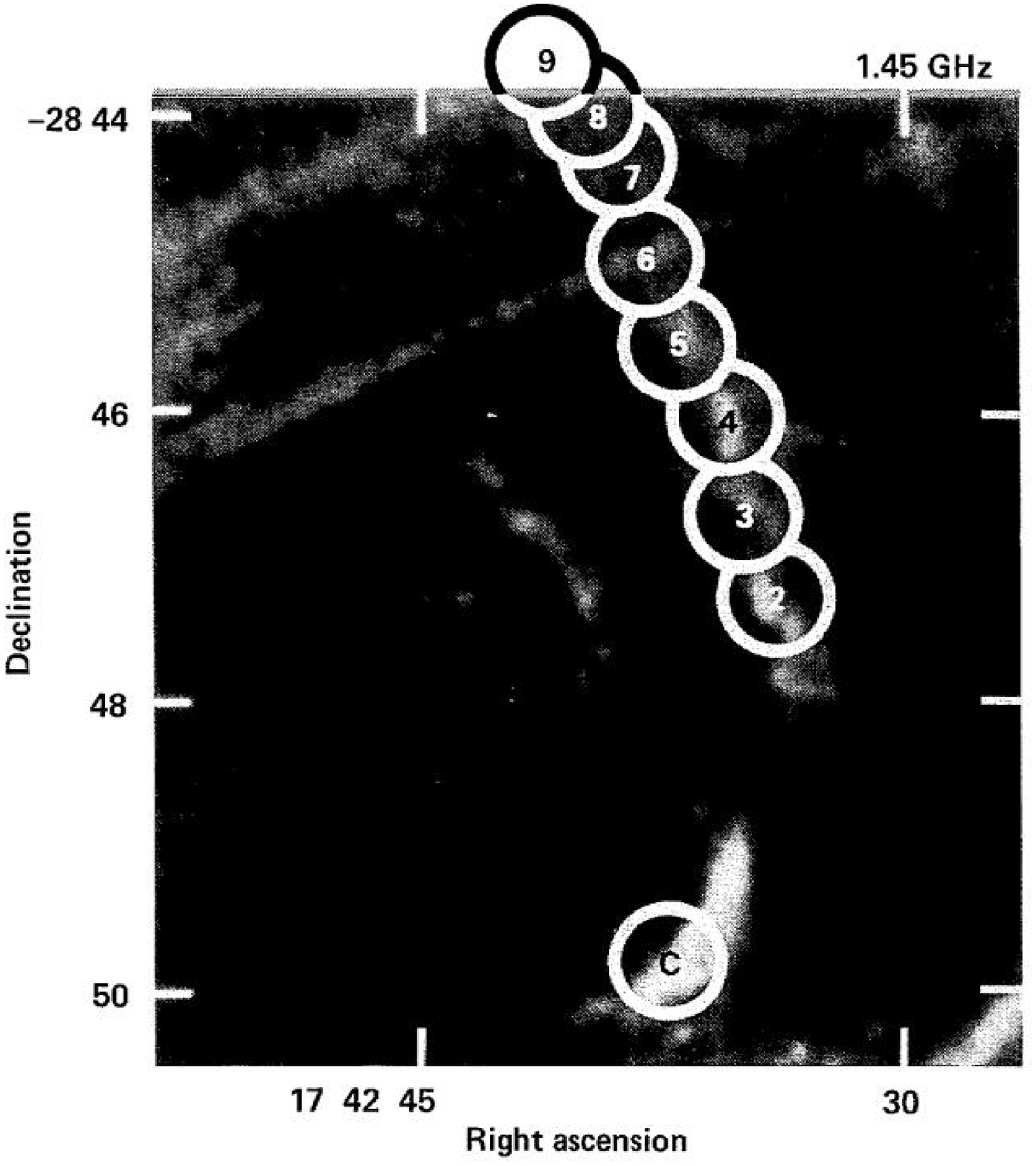}
\caption{Location of infrared observations overlaid on a VLA continuum radiograph 
(Yusef-Zadeh et al. 1984).The figure is taken from Erickson et al (1991, Fig. 1).
}
\end{center}
\end{figure}

 Radio VLA maps   show peculiar filamentary
structures $\sim$ 10' long located roughly 10' northeast of the Galactic center,
which suggest fragmentation of matter close to shock fronts.
The morphology and the radio polarization (Yusef-Zadeh, Morris, \& Chance 1984)
indicate that magnetic fields are important,
which is significant for models including shocks.
Moreover, it was found that the Arches Cluster is photoionizing the  region of straight and arched
filaments surrounding it  (E91).
This   led us to adopt composite models (shocks+ photoionization) 
to explain gas and dust spectra observed from    both thermal (arched)  and non-thermal (linear) structures.  

The clouds and filaments  at position G0.095+0.012 and the E2 thermal
"arched" radio filament near the Galactic center were observed by E91
in 8 positions (Fig. 6). 
To clarify the nature of these filaments, Erickson et al (1991, hereafter E91) have made FIR line 
and adjacent continuum
observations of [SIII]19 , [SIII]33, [OIII]52 and [OIII]88, [NIII]57, [SiII]35,
[OI]63, [OI]145, 
and [CII]158 from NASA's Kuiper Airborne Observatory (KAO). 
Upper limits were obtained for [SI]25, [FeII]26, and [NIII]36.

In this section 
the  modelling of  line and continuum spectra is presented.

\subsection{Position C}
\subsubsection{The line spectrum}

We have corrected the spectrum   from extinction  as indicated by E91.
In Table 3,  we compare our models with the observed corrected  line ratios normalized to [SIII]33.5 =1
 The best fit is obtained  adopting  \Ts=24,000 K and U=4 10$^{-3}$,
\Vs=65 \kms,  and \n0=200 \cm3. The magnetic field \B0= 2 10$^{-5}$ gauss
is similar to that  found  in S07 position 31.
The relative abundances which lead to the best fit of the observed spectrum, show that  C/H is  lower 
than solar by a factor of 1.65, while N/H is higher
by a factor of 1.5. Si, S, and Fe are  lower than solar  suggesting that they are trapped into grains.
Also in this case the clouds are  moving towards the hot source, i.e. the Arches Cluster.
 In our models, the temperature of the stars  (24000 K)  results phenomenologically
 because leading to the  consistent fit of all the lines.
E91 adopted a T$_{\it eff}$=35,000K atmosphere  from Kurucz (1979).
The LTE atmosphere has a very different UV SED from a black body (Simpson et al 2004, Fig. 6) 
so the entire modelling is different. 
 
\begin{table}
\caption{IR line spectrum at position C}
\begin{tabular}{lllll}\\ \hline  \hline
\      line    & obs$^1$      &I$_{\lambda}$/I$_{[SIII]33.5}$$^2$ &  m$_{C}$ \\ \hline
\ [SIII] 18.8 & 21.3$\pm$4.0 & 1.8         & 1.82 \\
\ [FeII] 26   & $<$9.4       &$<$0.2        &0.2\\
\ [SIII] 33.5 & 70.5 $\pm$1.1 & 1.         &1.\\
\ [SiII] 34.8 & 31.5 $\pm$1.2 & 0.42 & 0.41       \\
\ [NeIII] 36. & $<$0.7      &$<$0.009       & 0.009 \\
\ [OIII] 51.8 & 12.8$\pm$0.4  & 0.135      & 0.14 \\
\ [NIII] 57.3 & 15.7$\pm$0.5  & 0.159&       0.15 \\
\ [OI] 63.2   & 5.2$\pm$0.4   & 0.05        & 0.06 \\
\ [OIII] 88.4& 11.2$\pm$0.3  & 0.1        &0.08\\
\ [OI] 145.5 & 0.5$\pm$0.05 & 0.004 &        0.003 \\
\ [CII] 157.7 & 8.2$\pm$0.1 & 0.074 &      0.075 \\
\ \Hb (\erg) & -&- &       6.4e-5 \\
\  \Vs (\kms)& - &-   &65 \\
\ \n0 (\cm3) & - &-     &200  \\
\ \B0 (10$^{-3}$ gauss)& - &-    &0.02\\
\ \Ts (K)              & - & -&    2.4e4 \\
\  U                   &-& - &  4.e-3  \\
\ $D$ (10$^{14}$ cm)   & - &-    &9.7 \\
\  C/H &-&-&          2.0e-4 \\
\  N/H &-&-&          1.5e-4 \\
\  Si/H&-&-&         4.0e-6  \\
\  S/H &-&-&         1.0e-5 \\
\ Fe/H &-&-&          2.6e-6 \\

\hline
\end{tabular}

\flushleft

$^1$ 10$^{-18}$ W cm$^{-2}$

$^2$ extinction corrected (Erickson et al (1991, Table 1)
\end{table}

In Fig. 7 we show the profile of the electron temperature and
electron density,  and of the fractional abundance of
the most significant ions downstream as calculated by model m$_C$.
The model is matter bound.

\subsubsection{The continuum SED}

We try to constrain the model adopted  to  explain the line spectrum,   using the SED
of the continuum.
We plot in Fig. 8  the data from E91. The data cover
only the far-IR range,  but they are enough to show that with model
m$_{C}$ the continuum data are not reproduced.  In particular, the model dust reradiation peak  
is  shifted at a lower
frequency. We check whether  a higher \Vs could improve the agreement since
  higher \Vs  lead to  higher dust peak  frequencies.
We have adopted a rather large \Vs  compared with the radial
velocities  ($\sim$ 10 \kms)  measured by E91.
Morris \& Yusef-Zadeh (1989) suggest a mechanism to account for the ionization 
and radio emission based on a retrograde, high velocity of $\sim$ 100 \kms
cloud encountering the  poloidal magnetic field in the vicinity of the 
GC. Even with such a high velocity, the  dust peak  could not be  reproduced.

In relatively low shock-velocity regimes, a high flux dominates the ionization and heating
processes.  We have therefore
run a model with a very high U (=5), as we have done  for the dust peak relative
to the S07 observations. 
The other parameters are the same as those of model m$_{C}$. 
The fit to the IR data by the hot dust model is  satisfactory. 
Dust is not homogeneusly distributed  throughout the observed region.
Dilution of U can be explained by a larger distance from the  photoionizing source
  and/or by obscuring  matter between the radiation source and the emitting clouds.

\subsection{The  E2 arched filament}

 E91 reported the observations of the [SIII]33,
 [OIII]52, and [OIII]88
lines at eight positions along the E2 arched filament. They claim that the E2
filament is roughly tubular with a 10:1 length to diameter ratio. Moving northward
along the filament, the excitation decreases slowly and the line and continuum
brightness decrease by a factor of $\sim$ 2. 

In Table 4 we compare the calculated 
 [OIII]52/[SIII]33  and [OIII]52/[OIII]88  line ratios with the data
corrected for extinction. The lines observed are too few to fully constrain
the models. 
In Table  4 we refer to positions 2, 4, 6, and 8, where both the [OIII]52 and [OIII]88
lines are observed. 
We notice  by modelling that the line ratios  depend strongly on the preshock density.  
These ratios are significant because   [SIII]  refers to a ionization potential
lower than that of [OIII], so the trend of the [OIII]52/[SIII]33 ratio eventually  resembles
that of   the [OIII]/[OII] ratio,  assuming constant relative abundances.

\begin{table*}
\caption{IR line spectra in the E2 arched filament}
\begin{tabular}{cccccccccccccc}\\ \hline  \hline
\ line ratios  &\multicolumn{2}{c}{position 2} &\multicolumn{2}{c}{position 4} &\multicolumn{2}{c}{position 6} &\multicolumn{2}{c} {position 8} \\ \hline
\                   & obs & calc &  obs & calc & obs & calc & obs & calc \\
\  [OIII]52/[SIII]33 & 0.176 & 0.178&0.105&0.105 &0.0664&0.067&0.076&0.076\\
\  [OIII]88/[SIII]33 &0.116  & 0.112&0.079&0.079 &0.076&0.076&0.067&0.066\\
\  \n0 (\cm3)        &  -    &160   &-    &110   &-& 40 &    - & 80   \\
\hline
\end{tabular}

for all positions \Vs=65 kms, \B0= 2 10$^{-5}$ gauss  U=4 10$^{-3}$, \Ts= 24 000 K and the relative
abundances as for position C.
\end{table*}

\begin{figure}
\includegraphics[width=0.23\textwidth]{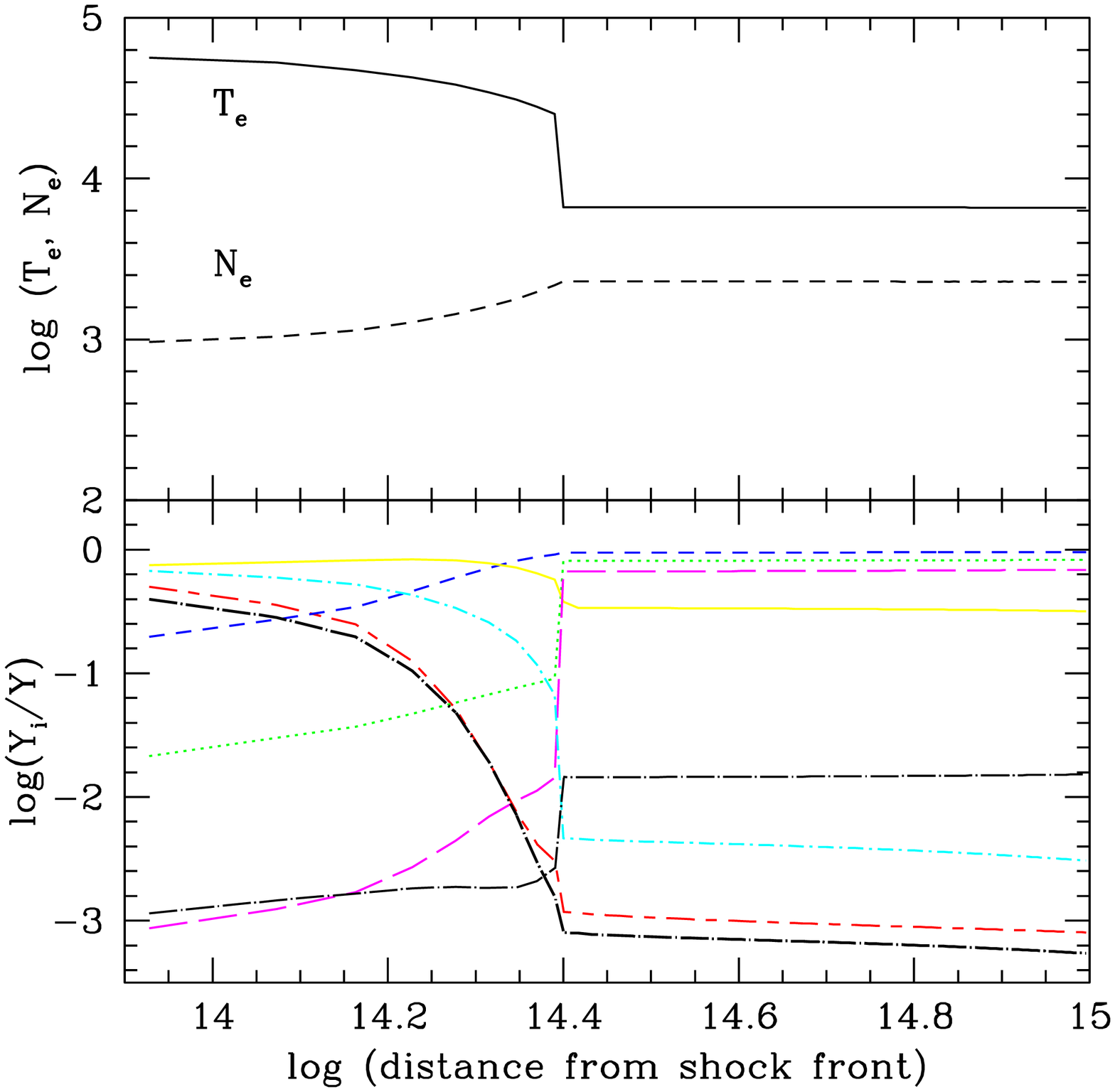}
\includegraphics[width=0.23\textwidth]{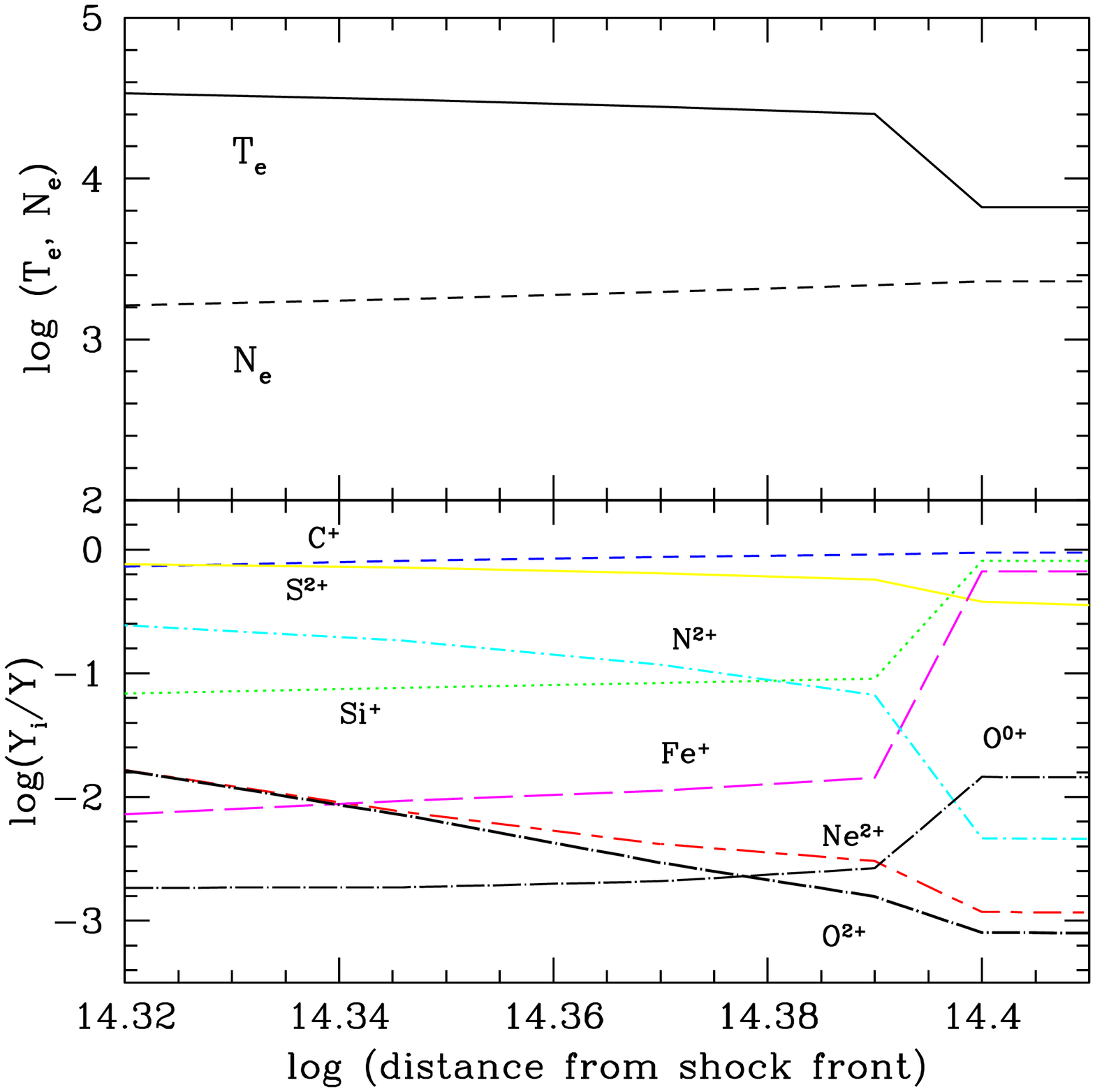}
\caption{Top : the profile of the electron temperature and  the electron density
downstream in position C, calculated by model m$_C$.
Bottom : the profile of the fractional abundance of the most significant ions.
The diagram on the right  presents a zoom  of the temperature drop region downstream.
}
\end{figure}

\begin{figure}
\begin{center}
\includegraphics[width=0.45\textwidth]{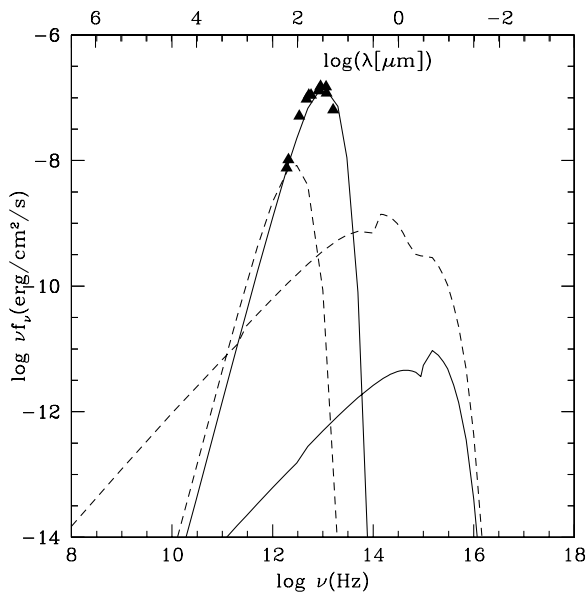}
\caption{The comparison of the calculated continuum SED in position C
with the  data (Erikson et al. 1991).
Short-dash : model m$_{C}$; solid : model calculated with
U=5 
 For all models two curves appear referring to the bremsstrahlung and to reradiation
by dust}

\end{center}
\end{figure}

\begin{figure}
\begin{center}
\includegraphics[width=0.45\textwidth]{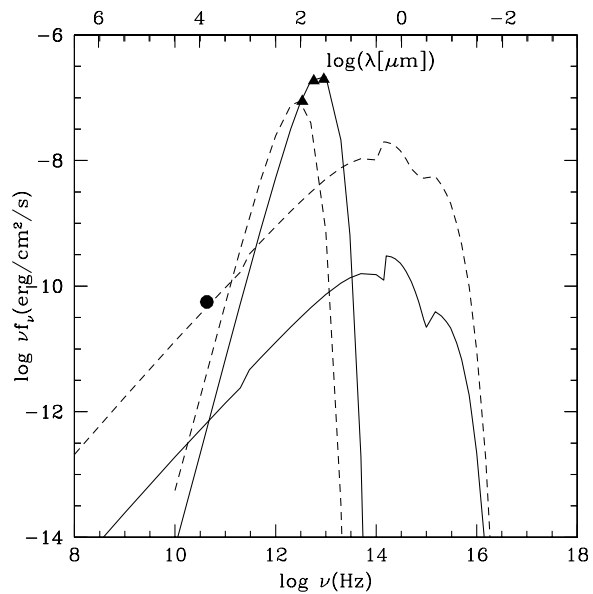}
\caption{The comparison of the calculated continuum SED 
with the  data in position 2 of the E2 strip (Erikson et al. 1991).
Short-dash : model calculated in position 2 (see Table 4); solid : model calculated with
U=5  For each model two curves appear, one refers to the bremsstrahlung,
the other peaking in the IR referring to the dust reprocessed radiation
}
\end{center}
\end{figure}

In Fig. 9 the continuum SED in position 2 is compared with the IR data
(E91, Table 2) corrected for extinction.
Using the model which leads to the best fit of the line ratios,
 dust  reaches a maximum temperature of $\sim$ 40 K,
while the data are better explained by a temperature of $\sim$ 94 K.
This relatively high temperature can be achieved by a very high U (see Sect. 3.3.1).
  The model
which  explains the line ratios is constrained by the datum in the radio range.
The contribution of the  high U cloud component in the line spectra
is  $<$ 5\%. 
The high U  clouds can be very close to  stars embedded within the filament, 
 or they are directly reached by radiation from the Arches Cluster stars,
 as previously explained.

Notice that  iron is highly depleted from the gaseous phase, therefore we can attribute IR radiation  to
iron rich grains (see Sect. 3.3.2).

\section{The spectra  in the  E2-W1 Crossing Strip}

\begin{figure}
\includegraphics[width=0.45\textwidth]{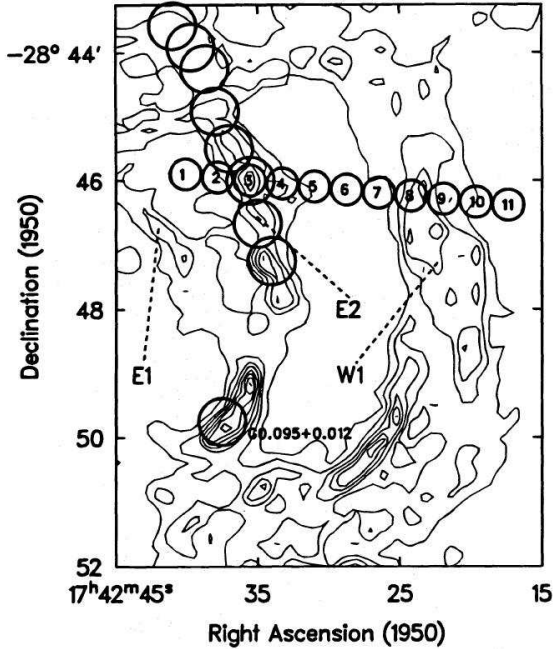}
\caption{Positions of infrared observations overlaid on a VLA 5 GHz
continuum map from the data of Morris \& Yusef-Zadeh (1989).
(Taken from Colgan et al. 1996, Fig. 2)
}
\end{figure}

The region  0$^o$.25 north of the GC   is characterised by $\geq$ 30 pc long, thin,
straight filaments of ionized gas which cross the Galactic plane
at right angles (Yusef-Zadeh et al. 1984, Simpson et al. 1997). 
 Their radio continuum emission is polarized and
 nonthermal, indicating 
relative strong  magnetic fields (e.g. Tsuboi et al 1985). 
In the north-western region, the linear filaments crossing the Galactic plane
intersect the Arched filaments, which emit thermal radio continuum.
It seems that the  linear and arched filaments are connected (Yusef-Zadeh \& Morris 1988).
The excitation mechanisms responsible for the emission for both sets of
filaments is controversial (Colgan et al 1996, hereafter C96, and references therein).
Photoionization from Sgr A West is excluded because  the photon flux is too low.
Photoionization by  a line of OB stars  close to the  filaments  is not suitable to  the
region's morphology. Collisional excitation of the lines, derived from the MHD model of
Morris \& Yusef-Zadeh (1989) is rejected on the basis of electron densities lower than
that of the adjacent molecular gas. Also embedded evolved stars not associated with the filaments
could provide some fraction of the continuum, however, Genzel et al (1990) claim that their 
luminosity is too low  to provide the infrared continuum.

It is now clear that the hot young star cluster (the Arches Cluster, Figer et al 1999) found
by Cotera et al (1996) and Nagata et al. (1995) is the main source of photoionization.
Moreover, the FWHM of the lines presented by Cotera et al. (2005) for the E1 filament
and the top of the E2 filament are relatively high and indicate that the effect of shocks 
is non negligible.

 C96, in their Table 1, present the far-IR line and continuum spectra 
in 11 positions of the strip between E2 and W1 thermal radio filaments in 
the Galactic Center "arch" (Fig. 10). 
In the following we will  try to explain the spectra by  composite models
that  were used previously in Sects. 3 and 4, namely, shock and photoionization
are  consistently adopted  to calculate  the line ratios. Comparison of calculated with observed
line ratios leads to the set of parameter which best describe the physical conditions in the
observed regions. We consider that the photoionizing radiation flux is provided by the stars in the
Arches Cluster.

The observations were made by the Kuiper Airborne Observatory (KAO) facility Cryogenic Grating Spectrometer (CGS)
(Erickson et al. 1985). 
The lines fluxes presented by C96, include [SIII]33, [SiII]35, [OIII]52, [OI]63, [OIII]88, and
[CII]158.

\subsection{Preliminary model investigation}

\begin{table*}
\begin{center}
\caption{The preliminary models}
\begin{tabular}{llllllll}\\ \hline  \hline
    model       &\Vs        & \n0     &  $U$  &   $D$ &  symbols  \\
                & (\kms)    & (\cm3)  & -   &  (10$^{15}$ cm)&  \\
  mp1  & 50        & 70      & 0.005&  1-10&dotted line linking  empty triangles (cyan) \\
  mp2   & 60        & 80      & 0.0005& 0.5-50& short-dashed  +  asterisks (5) (black) \\
  mp3   & 60        & 100     & 0.0005& 0.3-30&short-dashed  +  asterisks(7) (black) \\
  mp4  & 60        & 150     & 0.0005& 0.2-16& short-dashed  +  empty circles (black) \\
  mp5  & 50        & 70      & 0.001 & 1-2&dotted  +  empty pentagons (magenta)\\
  mp6$^1$   & 30        & 30      & 0.0015& 2-10&long-dashed  +  empty hexagons (green)\\
  mp7   & 50        & 60      & 0.002 & 10-100& solid  +  asterisks (3) (red)  \\
  mp8  & 50        & 70      & 0.002 & 0.8-14 & solid  +  asterisks (5) (blue) \\
  mp9  & 50        & 80     & 0.0015& 0.6-4.5& solid  +  dash (black)  \\
  mp10$^2$  & 60    &  60     & 0.002 & 30-150 & long dashed  +  asterisks (5)(red) \\
\hline
\end{tabular}

\flushleft

$^1$ m9  was  calculated adopting Si/H = 3.3 10$^{-6}$

$^2$ m10 was  calculated adopting \B0=10$^{-4}$ gauss

\end{center} 
\end{table*}

\begin{figure}
\includegraphics[width=0.42\textwidth]{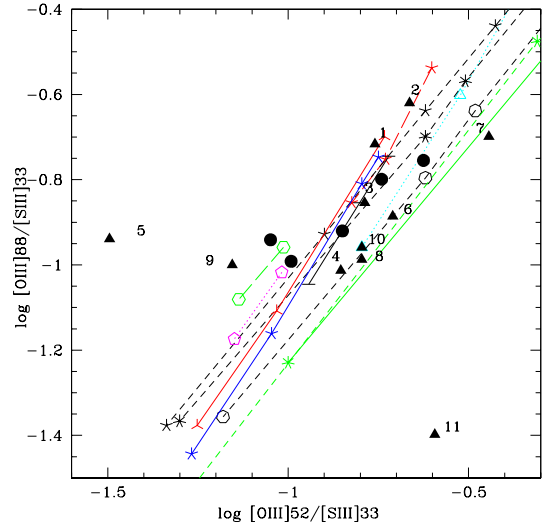}      
\includegraphics[width=0.42\textwidth]{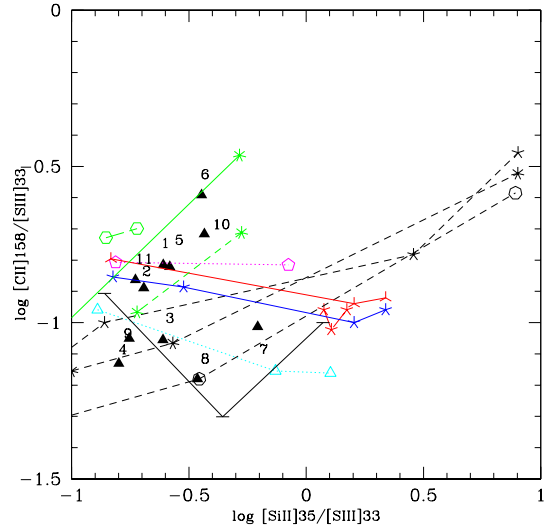}      
%includegraphics[width=0.42\textwidth]{mmilkye2_w1c.ps}
%includegraphics[width=0.42\textwidth]{mmilkye2_w1b.ps}
\caption{The comparison of model calculations with the  the most significant
line ratios  observed in the E2-W1 crossing strip. The numbers refer to positions
in Colgan et al. 1996, fig. 2. The observations are represented by black triangles.
 In the  top diagram we have  plotted  the data (black dots) from the observations 
 of  Erickson et al (1991, Table 2)
The models are described in Table 5 and in the text} .
\end{figure}

Before trying a detailed modelling of the line ratios, we  investigate the parameter ranges and 
their influence on  the different line ratios by comparing the observed line ratios  with a grid
of models in the  Fig.  11  plots.
The models (mp1-mp10) are  described in Table 5.  They are characterized by
the shock velocity \Vs, the pre-shock density \n0, the ionization parameter U,
and  a range of geometrical thickness of the clouds which are indicated by point symbols
throughout the model trends.
In this first trial we adopted  a relatively high \Ts (34000K) close to that indicated by E91.
For all the models   \B0=3 10$^{-5}$ gauss
except for m10 which was calculated by \B0=10$^{-4}$ gauss.  A stronger
magnetic field prevents the compression downstream which is generally regulated
by the shock velocity and the pre-shock density and leads to  unsuitable line ratios. 

On the top of Fig.  11,  [OIII]88/[SIII]33 versus [OIII]52/[SIII]33 is compared with
model results. The data from C96 (their Table 1) are shown as filled 
triangles.  
On the same diagram we have  plotted the data from the observations 
 of  Erickson et al (1991, Table 2)
from  G0.095+0.012 and the E2 thermal radio filament. The data are distributed
from left to right : 6, 8, 4, C, and 2.

This diagram is related to the physical conditions in the gas
emitting the lines from the second ionization level. It seems that both the
[OIII]88/[SIII]33  and [OIII]52/[SIII]33 ratios increase with increasing density,
 because the S$^{++}$ region downstream is more reduced than the O$^{++}$ one  at higher n. 
These ratios are 
 particularly correlated with the geometrical thickness of the filaments,  decreasing at higher $D$ .

To constrain the models, we   show  in Fig. 11 (bottom) the [CII]158/[SIII]33
vs. [SiII]35/[SIII]33  plot.
The two  spectra at positions 5 and 11 which were not reproduced by the models in the top diagram
are well included among the other positions in the bottom one. 
In fact, the  spectrum in position 5  shows an unreliable [OIII] 52 (0.5 $\pm $ 0.4) and that in position 11
 shows  an  unreliable [OIII] 88 (0.2 $\pm$ 0.3). In the  bottom diagram which
is independent from these lines, the two spectra  regain the "normal" trend.
We refrain from showing the error-bars in the diagrams for sake of clearness.

The spectra  at positions   6 and 10 are not reproduced by the grid of models
presented in Table 5. In fact, the relatively high \Ts maintains the gas ionized
to the second ionization level in a large region, leading particularly 
to underpredicted  [CII]/[SIII] line ratios. 
Cross-checking the results obtained by a detailed modelling of the data (Tables 6 and 7),   
models mc6 and mc10 (green lines, solid and short-dashed, respectively)  were plotted  on Fig. 11.
  Two main trends can be noticed in the bottom diagram.  In fact,
the combination of the input parameters leads to the stratification of the ions downstream
of the shock front, which is also reached by the photoionization flux from the stars (Figs. 3, 7, and 12).
For instance, a  relatively high \Ts and/or a high U maintain the gas ionized to a higher D 
(the distance from the shock front),
 while a higher n speeds up recombination  because the cooling rate  is $\propto$ n$^2$.
The shock velocity   yields  compression (increasing n) and  a relatively high temperature 
downstream ($\propto$V$_s^2$) leading to a characteristic stratification of the physical
conditions.
When \Ts and/or U are  relatively low and D  relatively large,  
the fractional abundance of S$^{++}$ is low  and the 
[SIII] line flux remains nearly constant at larger D throughout the filament.
On the other hand, the first ionization potential of Si  and C  (8.11 eV and 11.20 eV, respectively) are lower than
that of H (13.54 eV), so  Si and C remain   singly ionized at larger D. This leads to [CII]/[SIII]
 and [SiII]/[SIII] line ratios increasing with D.
When \Ts and /or U are relatively high (models mp1-mp10) and D  are  such that  S$^{++}$ and C$^+$ fractional abundance
are still increasing, [CII]/[SIII]  slightly decreases.  As soon as D reaches the S$^{++}$ recombination distance,
[CII]/[SIII] increases.
[SiII/[SIII]  has an increasing trend because of its  very low ionization potential.

\begin{figure*}
\centering
\includegraphics[width=0.23\textwidth]{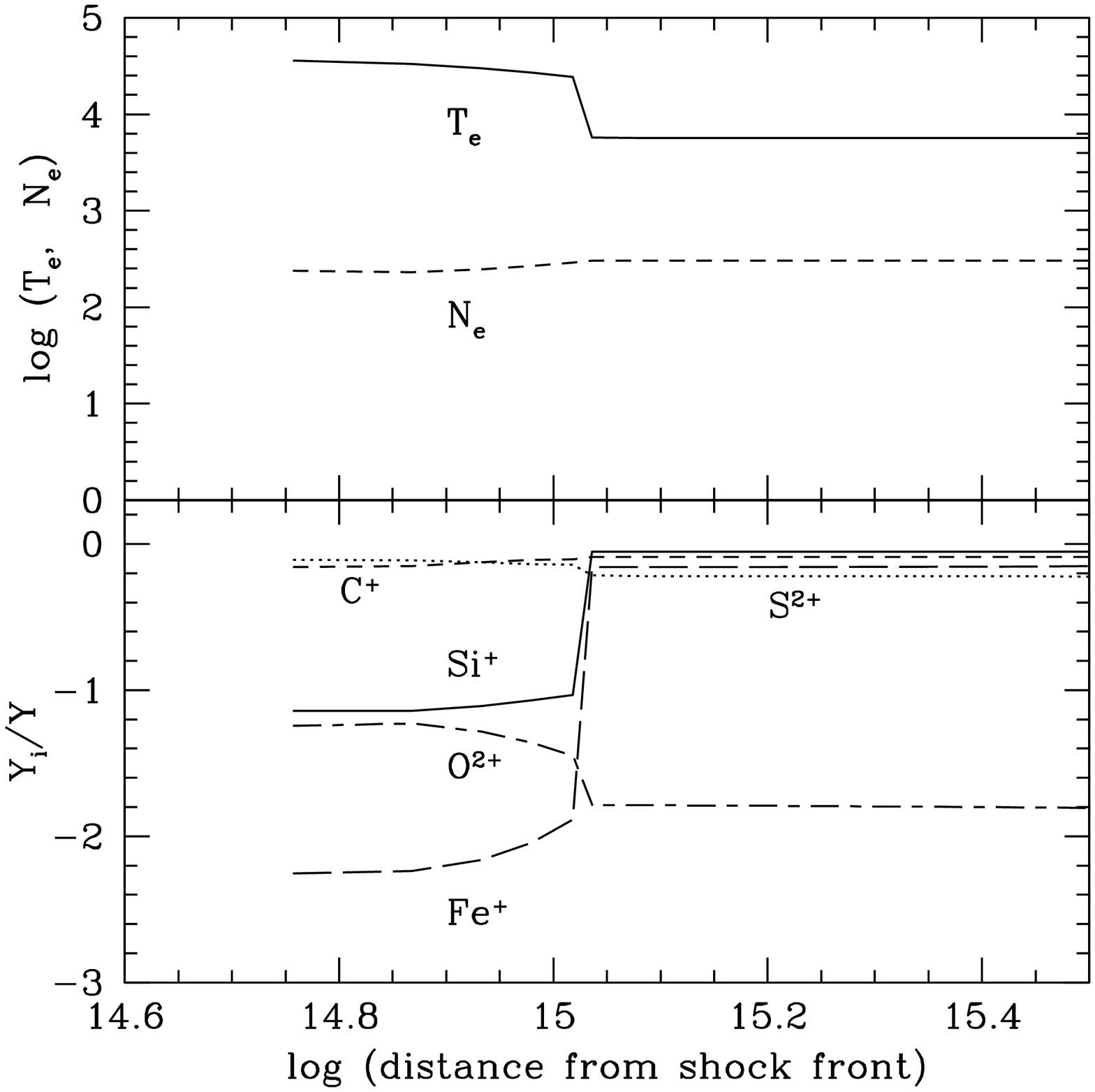}    
\includegraphics[width=0.23\textwidth]{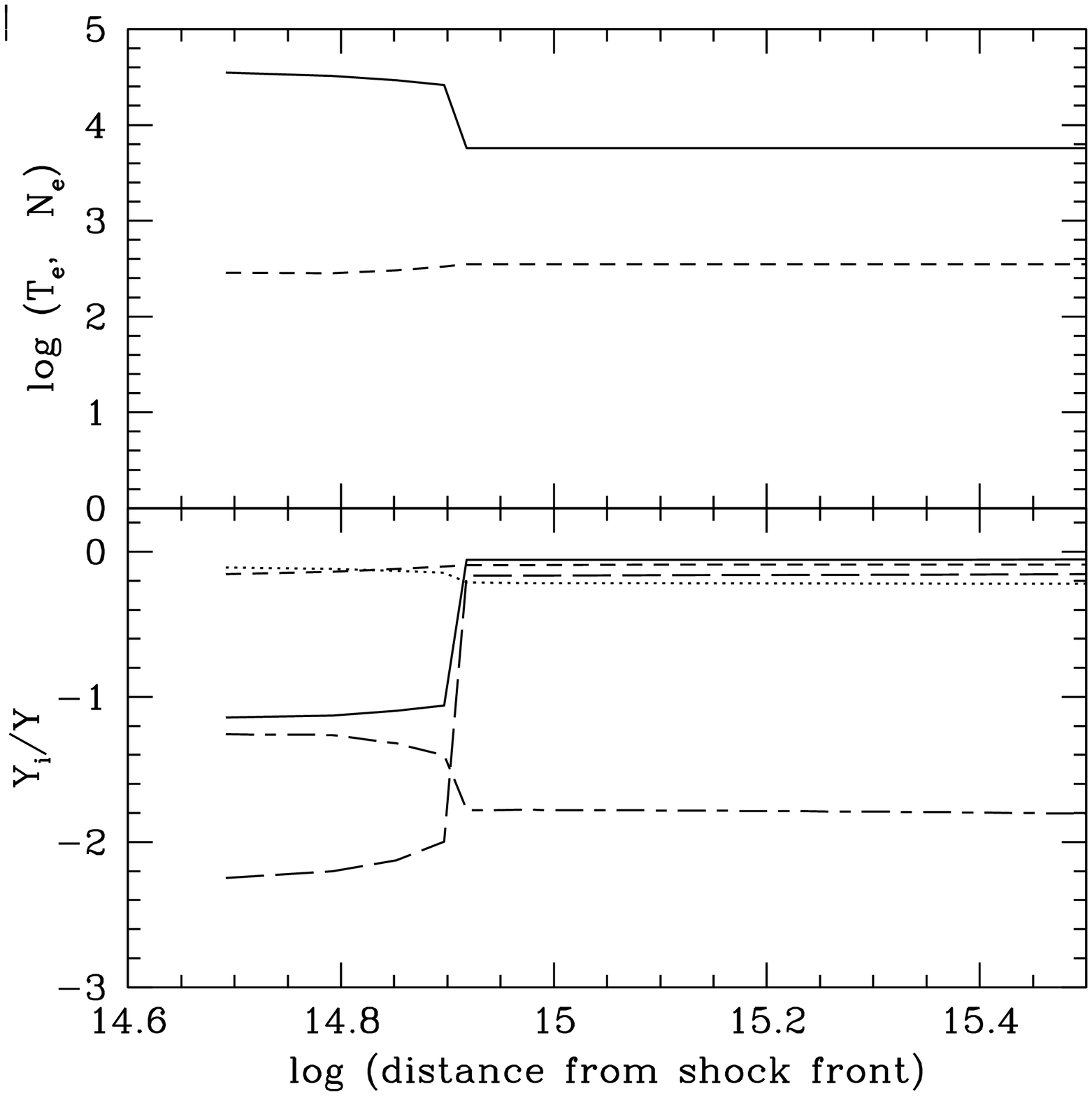}    
\includegraphics[width=0.23\textwidth]{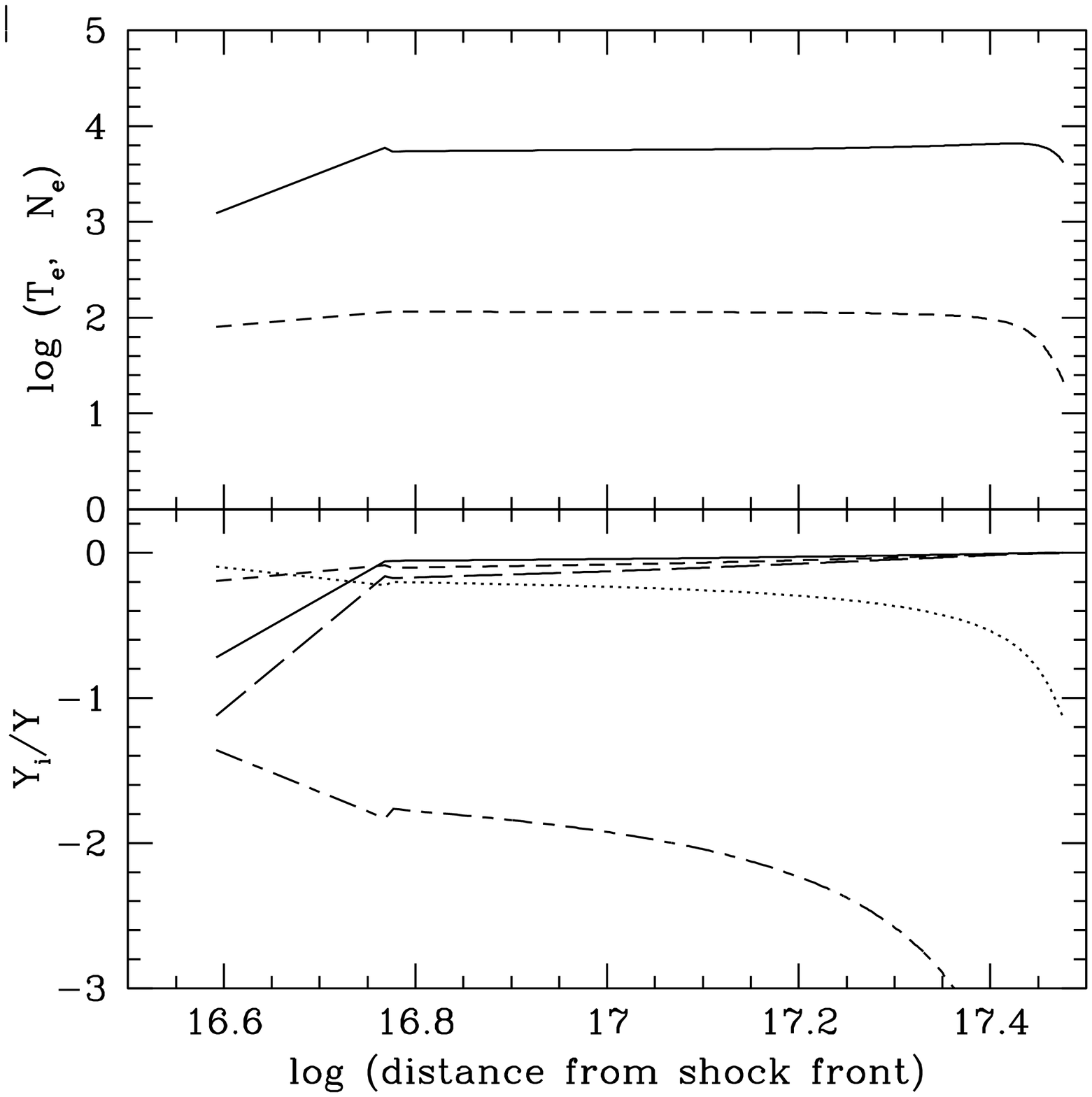}    
\includegraphics[width=0.23\textwidth]{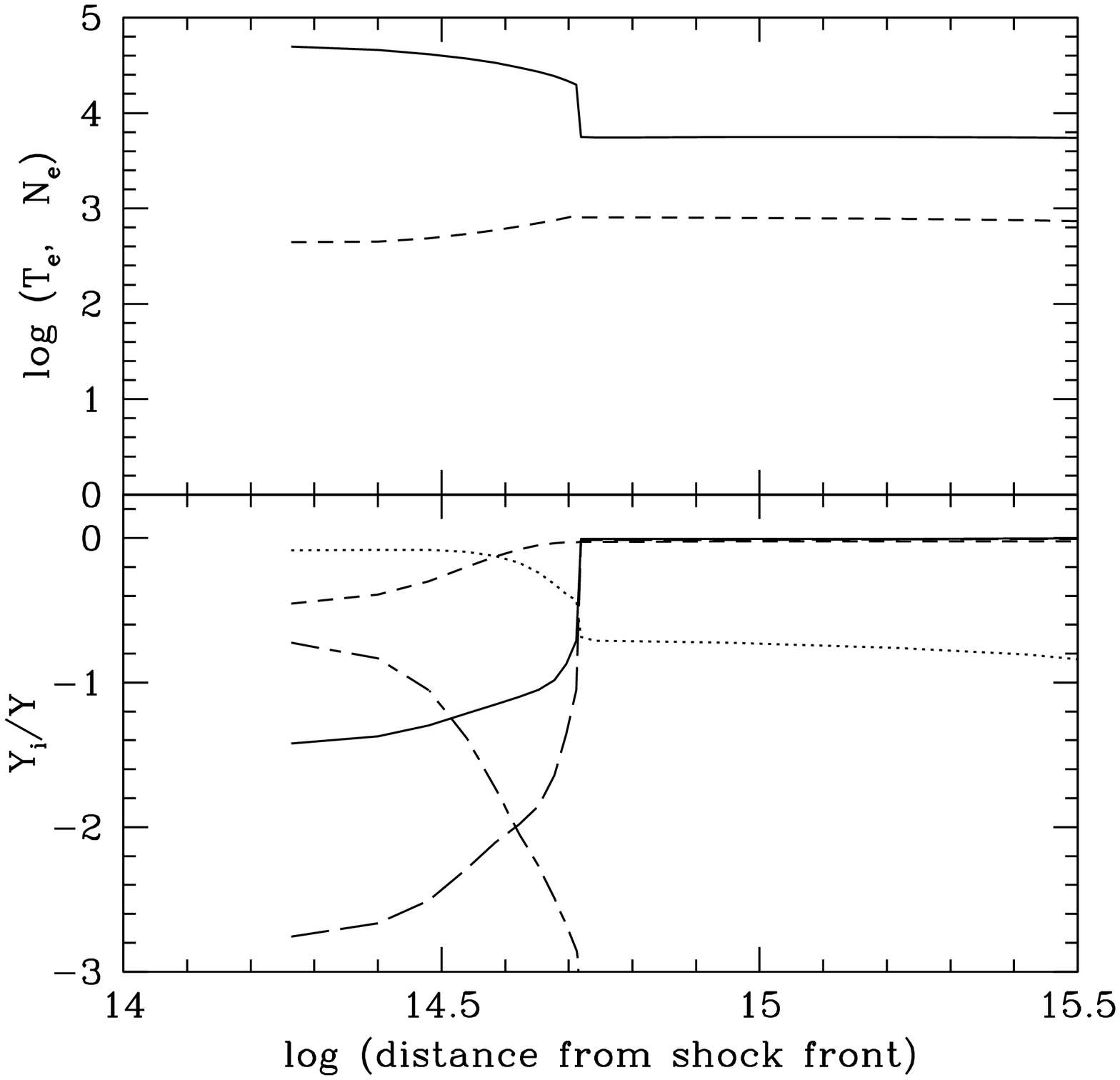}    
\caption{ The profile of the electron temperature and of the electron density
(top  diagrams) and the distribution of the fractional of the most significant ions
(bottom diagrams) throughout a cloud corresponding to models (Table 5) mp7, mp8, mp6, and mp3,
from left to right, respectively}
\end{figure*}

To better understand the trend of the models presented in Table 5,
 the profiles of the electron temperature, of the electron density,
and of the fractional abundance of the  more significant ions 
downstream are  shown in Fig. 12 for models mp7, mp8,
mp6 and mp3 from left to right, respectively.
Compression is relatively small for \Vs $<$ 100 \kms.
The best fitting models are matter bound as can be guessed from the relatively
low [SiII]/[SIII] line ratios.

Summarizing, the trend of the data in the E2-W1 strip was recovered using models with  \Ts = 27000K.
We conclude that, also on the basis of the conditions   found  in position C (\Ts =  24000K) 
a relatively low temperature is more suitable to the stars  close to  the observed positions.

\subsection{Detailed modelling}

The  absolute  flux of the [SIII]33 line
 is the strongest one so we will  consider line ratios to [SIII]. 
The  ratios of the observed corrected  lines fluxes to [SIII]33 
are  reported in Table 6. 
The line  fluxes were corrected according to C96 factors.
The results of modelling
are given in the row below that containing the data,  for all positions.
The  selected models numerated from mc1 to mc11  are described   in Table  7.

\begin{table*}
\caption{Comparison of observed (corrected) with calculated IR line ratios} 
\begin{tabular}{ccccccccc}\\ \hline  \hline
\ Position   & [SIII]33& [SiII]35 & [OIII]52 & [OI]63 & [OIII]88 & [CII]158 \\
\    1&  10.&  2.46&  1.29&  0.15&  1.26&  1.53\\
\  mc1&  10   &  2.66&  1.3 &  0.2 &  1.26&  1.44\\
\    2&  10.&  2.03&  1.61&  0.00&  1.58&  1.29\\
\  mc2&  10.  &  2.1 &  1.63&  0.1 &  1.58&  1.26\\
\    3&  10.&  2.45&  1.21&  0.15&  0.94&  0.88\\
\  mc3&  10.  &  2.4 &  1.28&  0.14&  0.93&  0.86\\
\    4&  10.&  1.59&  1.04&  0.32&  0.64&  0.74\\
\  mc4&  10.  &  1.59&  1.03&  0.31&  0.66&  0.79\\
\    5&  10.&  2.62&  0.24&  0.44&  0.76&  1.51\\
\  mc5&  10.  &  2.77&  0.46&  0.43&  0.76&  2.36\\
\    6&  10.&  3.58&  1.44& -0.45&  0.85&  2.56\\
\  mc6&  10.  &  3.8 &  1.44&  0.42&  0.85&  2.5\\
\    7&  10.&  6.21&  2.65&  1.06&  1.34&  0.97\\
\  mc7&  10.  &  6.4 &  2.54&  0.96&  1.4 &  1.06\\
\    8&  10.&  3.44&  1.19& -0.08&  0.68&  0.66\\
\  mc8&  10.  &  3.2 &  1.2 &  0.085& 0.67&  0.6\\
\    9&  10.&  1.76&  0.52&  0.10&  0.66&  0.89\\
\  mc9&  10.  &  1.73&  0.52&  0.10&  0.66&  0.9\\
\   10&  10.&  3.68&  1.19&  0.46&  0.74&  1.92\\
\  mc10& 10.  &  3.5 &  1.17&  0.47&  0.73&  1.4\\
\   11&  10.&  1.87&  1.89&  1.22&  0.26&  1.37\\
\  mc11& 10.  &  2.0 &  1.6 &  1.29&0.8   &  1.38\\
\hline
\end{tabular}
\end{table*}

\begin{table*}
\caption{The models adopted in the E2-W1 strip}
\begin{tabular}{lllllllllll}\\ \hline  \hline
\  model & \Vs&   \n0&    \B0   &  U     &  Si/H  &   S/H   &  C/H   & D   \\  \hline
\    mc1&   65&    91&     6.e-5&   2.e-3&  4.e-6 &   1.3e-5&  3.5e-4&2.9e15\\
\    mc2&   65&    91&     6.e-5&   2.5e-3& 4.e-6 &   1.3e-5&  3.5e-4&2.3e15\\
\    mc3&   65&   150&    7.e-5 &   3.0e-3& 5.e-6 &   3.5e-5&  3.5e-4&1.7e15\\
\    mc4&   65&   170&    5.e-5 &   2.3e-3& 2.7e-6&   1.5e-5&  3.3e-4&1.35e15\\
\    mc5&   73&   10 &    8.e-5 &   8.e-4 & 5.e-6 &   4.e-5 &  9.e-5 &2.7e18\\
\    mc6&   75&   190&    5.e-5 &  1.2e-3&  1.6e-6&   1.e-5 &  3.3e-4&2.9e15\\
\    mc7&   70&   280&    5.e-5 &  1.9e-3&  6.e-6 &   1.e-5 &  2.8e-4&9.e14\\
\    mc8&   70&   150&    5.e-5 &  7.e-3 &  9.e-6 &   1.2e-5&  3.4e-4&1.9e15 \\
\    mc9&   70&   30 &    5.e-5 &  3.e-3 &  2.3e-6&   1.2e-5&  9.e-5 &7.3e16\\
\   mc10&   70&   140&    3.e-5 &  3.5e-3&  3.9e-6&   1.e-5 &  4.e-4 &1.6e15\\
\   mc11&   70&   200&    1.e-5 &  3.1e-3&  1.8e-6&   1.e-5 &  3.6e-4&9.0e14\\
\hline
\end{tabular}

For all models \Ts=2.7 10$^4$ K is adopted
\end{table*}

The modelling is constrained by the [OIII]/[OI] line ratio, which depends strongly on the ionization
parameter, while the [OIII]52/[OIII]88 ratio depends on the density.
The shock velocity is not strongly affecting the line ratios, because  lines from relatively high ionization
levels were not reported. 
So we have chosen, as a first guess, \Vs in the range of the shock velocities explaining the spectra
observed by S07 (Table 2) in the region between E2 and W1.  The ranges of the other  parameters, \n0, \B0, and U,
were suggested by the preliminary investigation (Sect. 5.1) which also  leads to relatively low \Ts.
 
All the results presented in Table 6 were consistently calculated. In position 5 the density 
adopted to explain
 the very low [OIII]52/[OIII]88 line ratio is  exceptionally low. This was already   realized by C96.
Even  by \n0=10 \cm3, the calculated value is  lower than the observed one. Notice, however, that
the error given by C96 in their Table 1 for the [OIII]52 observed line flux is $\sim$ 80 \%.

The results are shown in Fig. 13. In  Fig. 13a the parameters depending on the shock are given
as a function of position.
The preshock density shows two deep minima, at positions 5 and 9. As  expected, the shock velocity  has a maximum
in position 5 denoting  that the shock velocity range is  relatively large.
The pre-shock magnetic field shows a decreasing trend in agreement with the results for \B0 obtained by explaining
 S07 observations between  positions 31 and 34.
In  Fig. 11a the [OIII]88/[OIII]52 ratio shows a  profile similar to that of the density, while the radio
distribution taken from C96 and shown in the  top panel, is not well correlated with the density. 
This can be explained by recalling that  
radio  and line emissions occur from different regions of the gas downstream in each cloud.

Interestingly, the distribution of both the radio fluxes at 43 and 1.4 GHz shows
that the two maxima do not correspond exactly  to the maxima in  D (Table 6) of 2.7 10$^{18}$ cm in position 5
and 7.3 10$^{16}$ cm in position 9.

In Fig. 13b, the profile of U is shown and compared  with the observed [OIII]52/[OI]63  and
[SiII]/[CII] line ratios (top panel). The oxygen line ratio follows the trend of U. The two U minima  in positions
5 and 9 correspond to the minima in \n0. The opposite would be expected because the ionization
parameter is reduced by crossing regions of dense gas. 
This indicates that the origin  of the U minima in both positions 5 and 9, is different. 

In  Fig. 13b (bottom panel) the relative abundances are shown for Si, S, and C, showing
that carbon is depleted from the gaseous phase at positions 5 and 9. Then, carbon grains, most probably
PAH,   screen the flux from the hot source and the  radiation flux   does not lead to 
full evaporation of the grains.
Also, Si is   trapped into dust grains
because  it shows depletion from the gaseous phase along all the strip.

The  other relative abundances adopted by the models
 are N/H = 10$^{-4}$, O/H = 6 10$^{-4}$, and Ne/H = 10$^{-4}$.

\begin{figure}
\begin{center}
\includegraphics[width=0.45\textwidth]{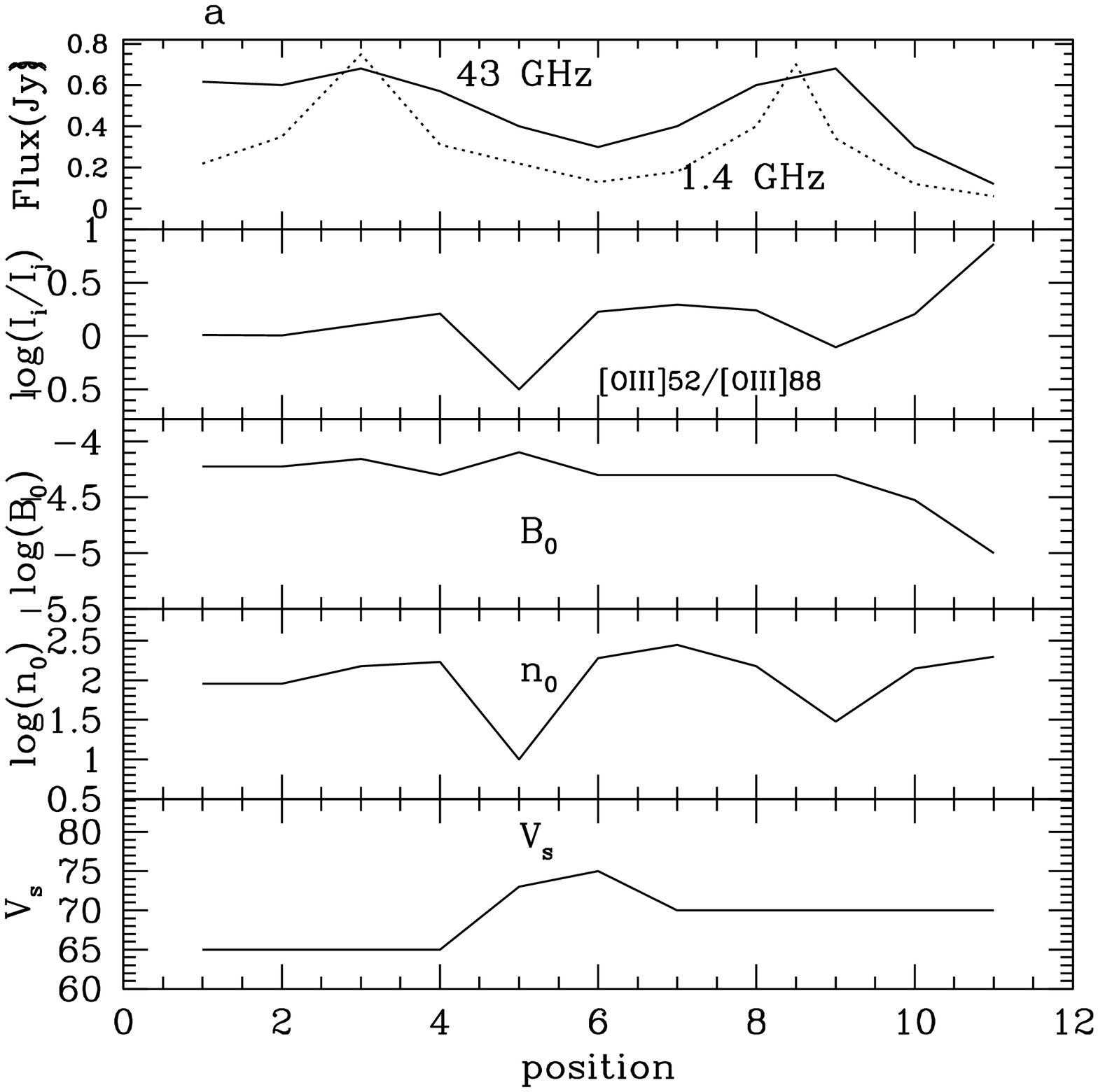}
\includegraphics[width=0.45\textwidth]{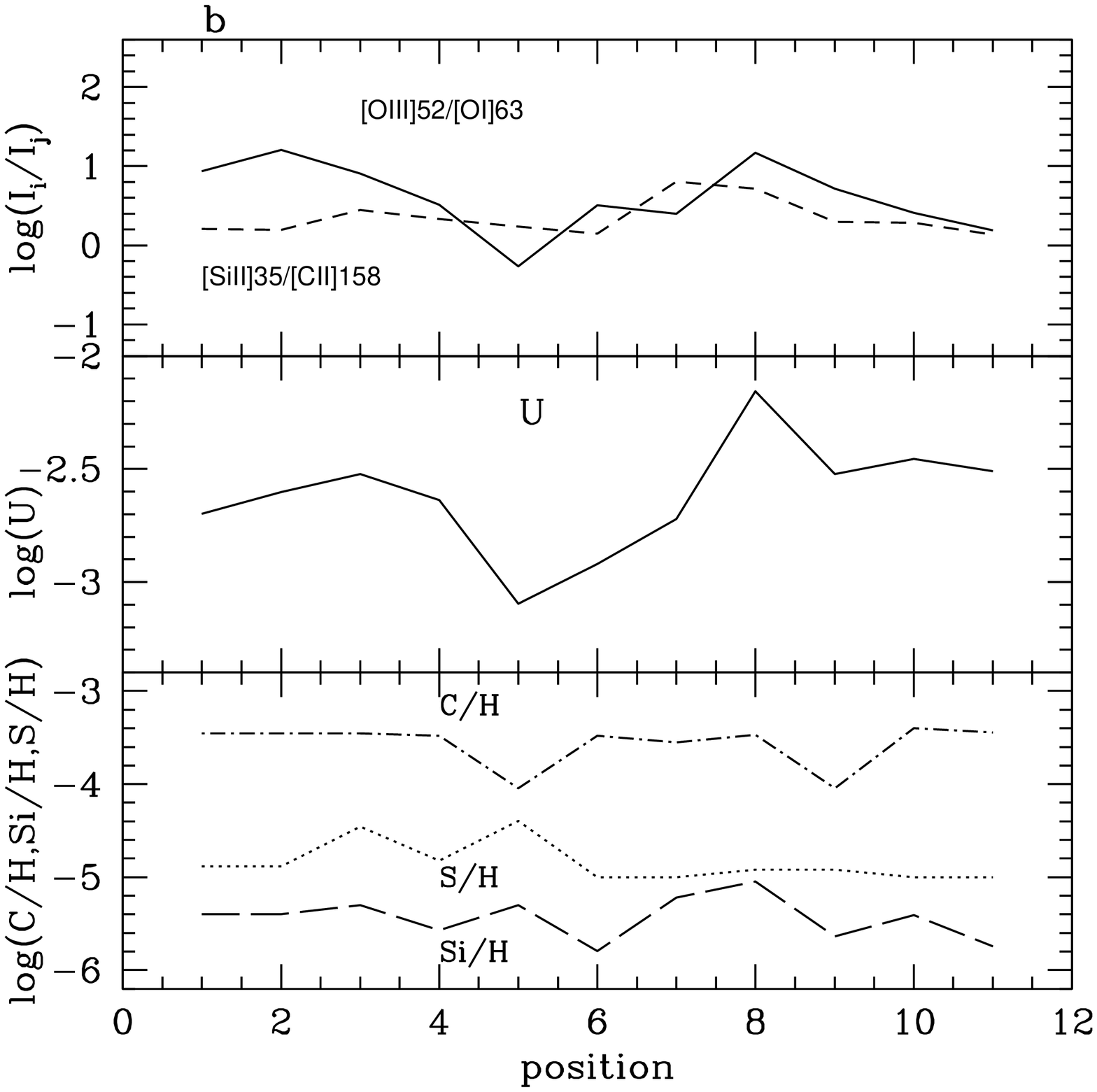}
\caption{The  results
along  the different positions  in the E2-W1 crossing strip.
a :the parameters depending on the shock.
b : the parameters depending on photoionization and the relative abundances
}
\end{center}
\end{figure}

In Fig. 14a  we   present the  continuum  SED in position 2.
The diagram on the left shows the modelling of the data.
The short-dashed lines show the bremsstrahlung and dust reradiation
calculated by  model mc2 which best fit the line ratios.
The data in the IR (C96, Table 1) are not well reproduced by the model
and indicate that dust grains are heated to a higher temperature,
because the  dust reradiation maximum  is shifted towards higher frequencies.
This  result was  found previously modelling the  data by S07 and E91. 
The best fit is obtained increasing U by a factor of 2000
and reducing \agr to 0.01 \mum. This leads to a maximum dust grain temperature of 88 K.
In order to reduce  the contribution
of such a dusty cloud to the line spectra, a $d/g$ =0.4  by  mass is adopted.
Grains are neither  destroyed by  evaporation  nor sputtered because the stars are not hot enough
and the shock velocities  are relatively low, respectively.
It seems that these grains  are  not explained by PAHs because C is not depleted from the gaseous phase 
in position 2. They could be explained by  eroded silicates and/or iron species.

The slope of the radio continuum is an interesting issue.
In fact,  the non-thermal or thermal character of the  emitting clouds is
determined on the basis of radio observations. The non-thermal character of the radio emission should confirm the
presence of shocks. Synchrotron radiation, created by the Fermi mechanism at the shock front, is observed in the
radio range from nebulae ionised and heated by  radiation from both the stars and shocks.
The relative importance of synchrotron radiation to bremsstrahlung determines the non-thermal or thermal character.
Fig. 14a shows that the radio datum at 43 GHz can be explained by thermal bremsstrahlung
as well as by dust reradiation. If it  corresponds to dust reradiation, 
 the synchrotron radiation flux created 
 could also contribute. We do not have enough data  in the radio range to confirm this.
On the other hand, if the radio flux which refers to the data  from Sofue et al. (1987) follows the bremsstrahlung
trend, it can  indicate some free-free self-absorption towards lower frequencies.
For comparison, we have added in Fig.  12a and b  the synchrotron power-law  radiation flux  (long-dashed line) which
 clearly follows a different trend.

To investigate the continuum   for the other positions, we  show in  Fig. 14b the data in both the IR
and  radio frequency ranges  for all the  positions.
The  results found for position 2  are  valid on a large scale also  for the other positions.
The dust temperatures are now constrained by radio data at 43 GHz. We have thus  added  the black body
flux corresponding to 200 K (black dotted line).
Such  a dust temperature  is easily reached by small grains.

In  Fig. 14c, a zoom on the dust reradiation maximum is shown. We can conclude
that dust cannot reach temperatures higher than 200 K.
In most positions there is  an absorption feature at wavelengths $\geq$ 30 \mum.
Even if  the data  are  insufficient to  determine the absorption and emission bands of typical grains,
we suggest that the  feature at $\sim$ 30 \mum is not so rare, since it was discovered
from ground based observations in 63 Galactic objects : 36 evolved carbon stars, 14 PPNe, and 13 PNe
(Hony et al. 2002). In our Galaxy, this feature, whose carrier seems to be MgS species,
 occurs from extreme AGB stars on to later stages (Jiang et al. 2008).

\begin{figure*}
\includegraphics[width=0.33\textwidth]{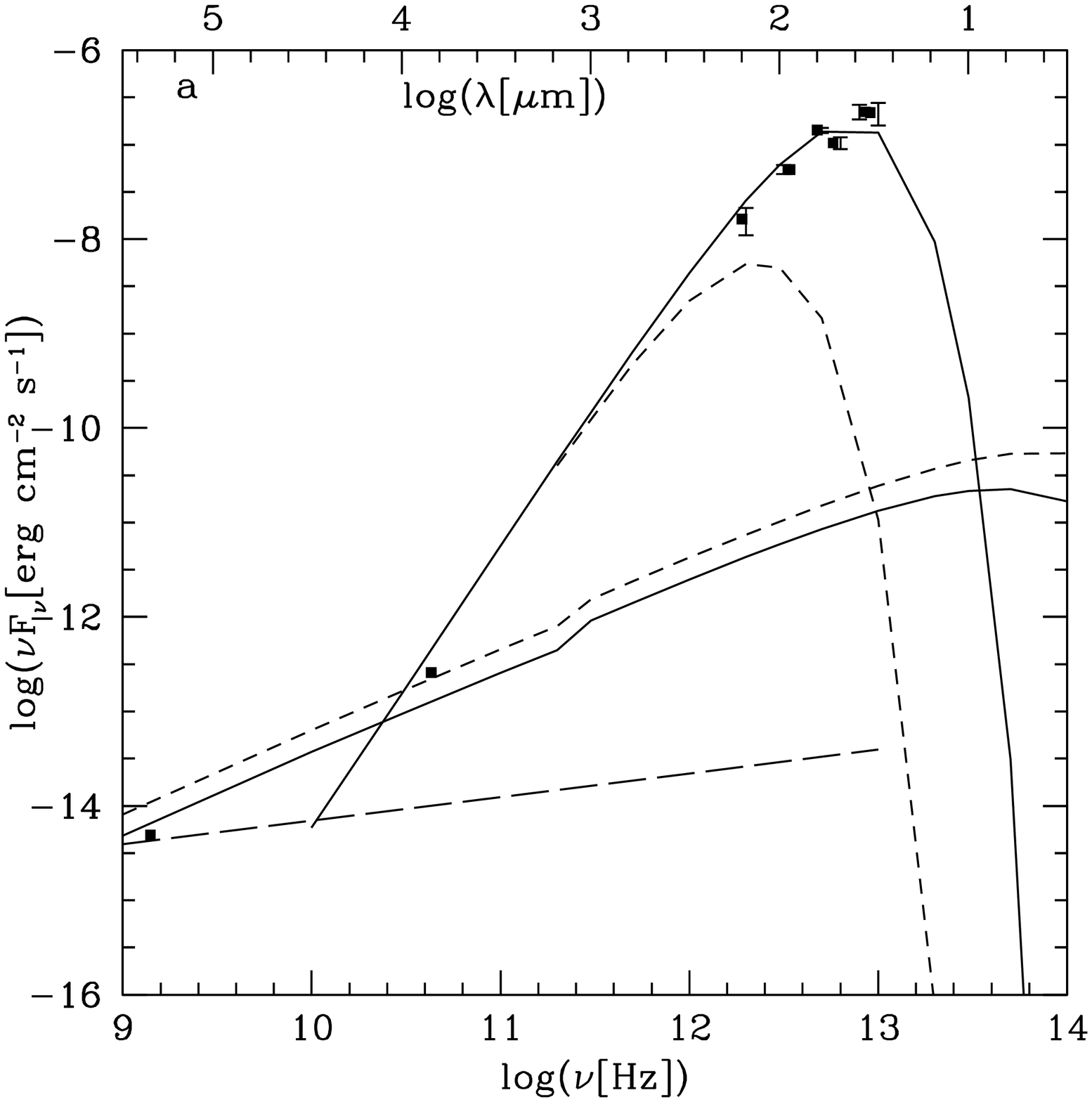}
\includegraphics[width=0.33\textwidth]{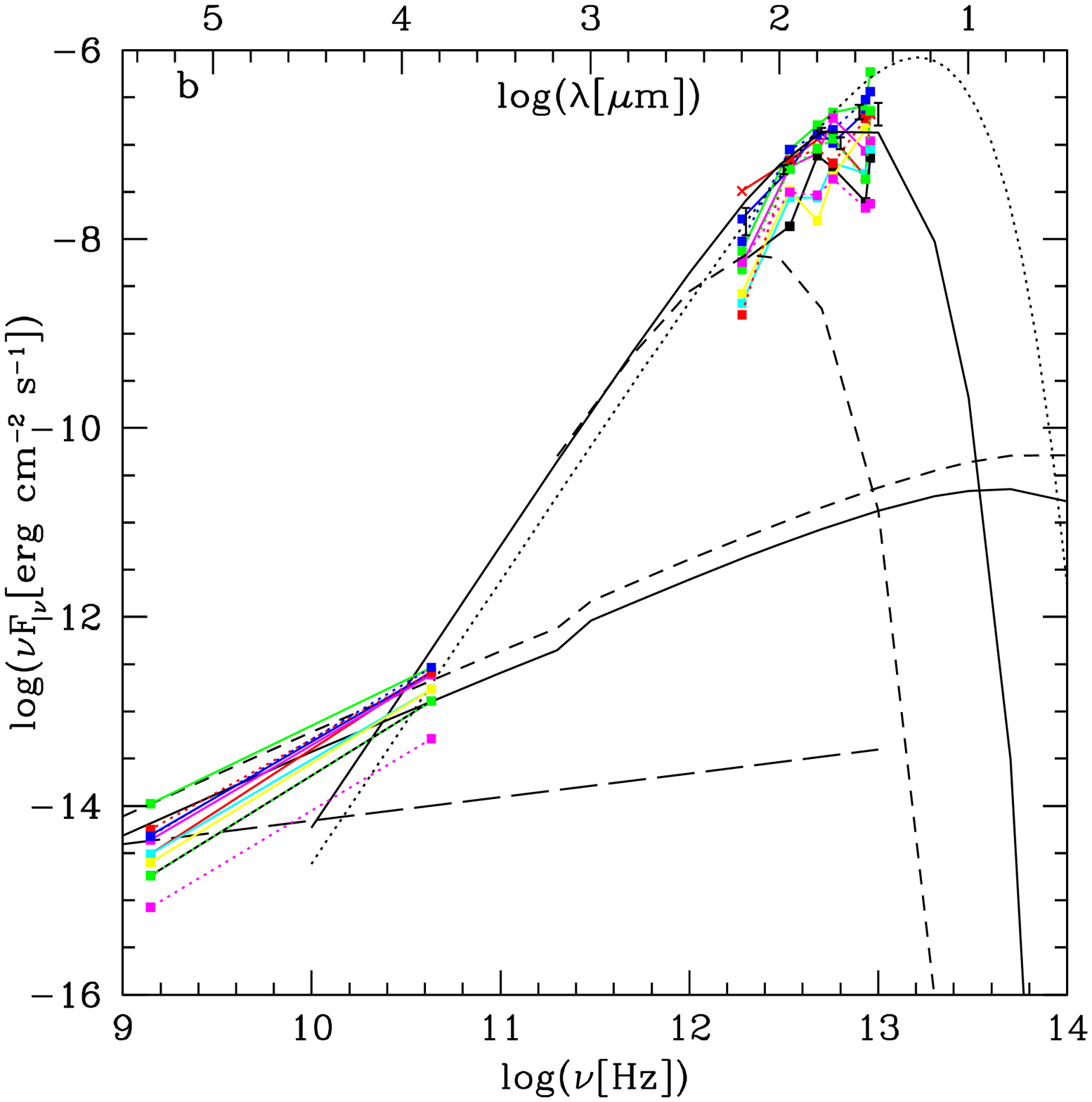}
\includegraphics[width=0.33\textwidth]{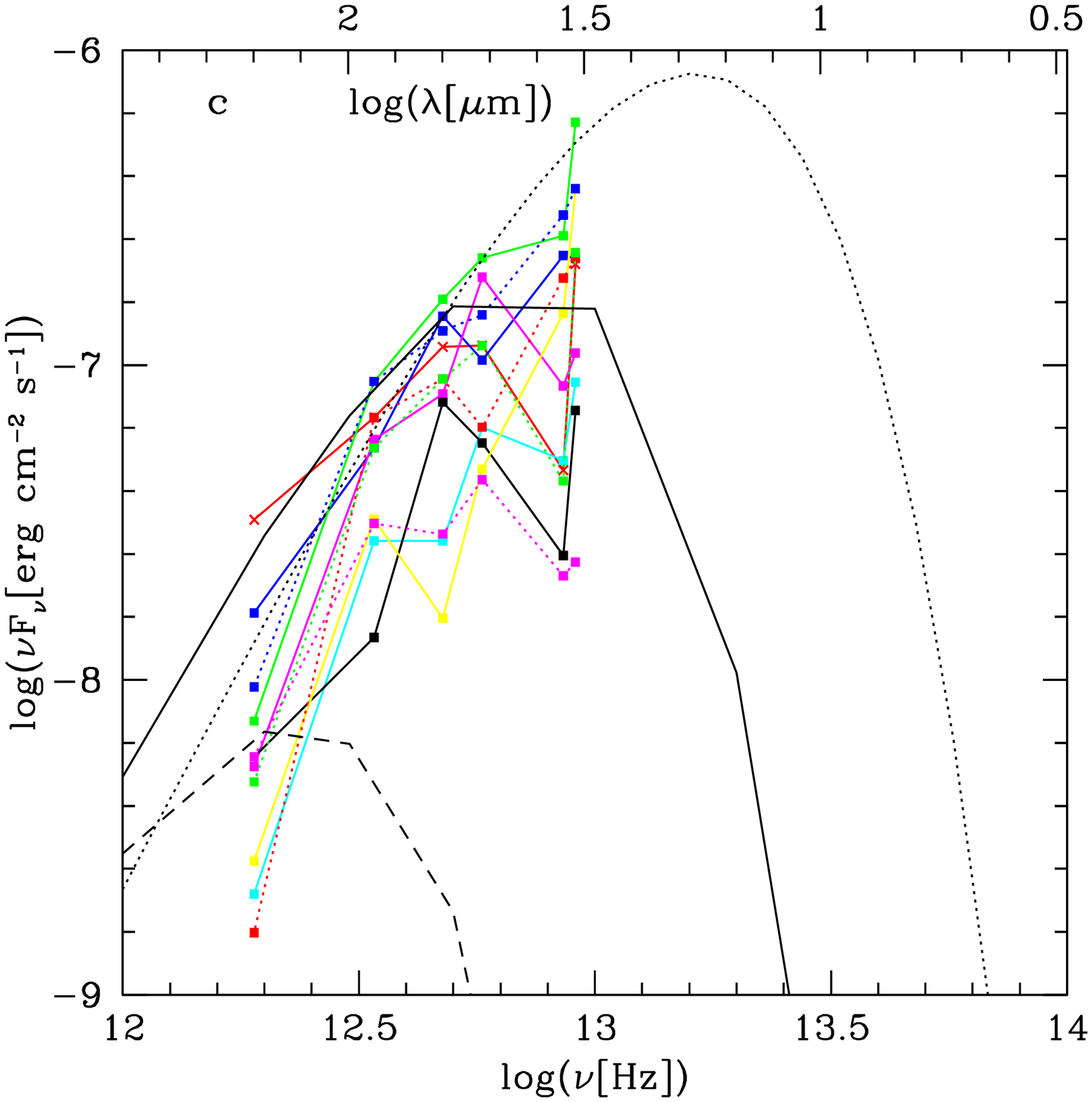}
\caption{The continuum SEDs in the E2-W1 Strip.
a : position 2. short-dashed : model mc2;
long-dashed : synchrotron radiation; solid line : model
calculated with U=5 and \agr=0.01 \mum.
b : the comparison for all position. Solid lines :
red : 1; blue : 2; green : 3; magenta : 4; cyan : 5; black  : 6;
yellow : 7 . Dotted lines :  red : 8; blue : 9; green : 10; magenta : 11.
c : a zoom in the IR maximum.
 For all models two curves appear referring to the bremsstrahlung and to reradiation
by dust}

\end{figure*}

\section{Concluding remarks}

We have modelled the IR spectra observed in the region near the Galactic center
with the aim of   determining the physical conditions in the different
observed positions. We have obtained  the results  by  comparing
 the calculated  with the observed line  ratios and  of the continuum  SED data.
Our models  account for the coupled effect of the shocks and the photoionizing flux.

The models  are matter-bound, indicating a high degree of fragmentation of matter, that
 is characteristic of turbulent regimes.
 We have found that  the shocks propagate  towards the photoionizing source,
i.e. the star clusters,  suggesting that gravitation  may prevail over 
eventual wind effects. 
The shock velocities range between $\sim$ 65 and 80 \kms. Indeed, they are not high enough
to produce X-ray emission by bremsstrahlung (e.g. Fig. 5) and  to collisionally heat the
dust grains to the observed temperatures of $\sim$ 150 K in some positions of the Arched Filament
region and of $\sim$ 88 K in the E2-W1 crossing strip.
In the downstream regions,  the characteristic  electron temperature and density 
profiles lead to  a good agreement of calculated line ratios from different ionization levels, 
with the observed ones.

The results obtained with pure photoionization models which account  on the [OIII] and 
[SIII] line ratios (e.g. Rodr\'{i}guez-Fern\'{a}ndez et al. 2001, Simpson et al. 2007, etc) demonstrate 
that photoionization from the clusters  affects the intermediate ionization level line ratios. 
However, adopting the composite models, detailed results can be found also for the shock velocities,
pre-shock densities, and pre-shock magnetic fields, by  modelling different level lines.

The pre-shock densities range between $\geq$  1 \cm3 in the ISM  and $\geq$ 200 \cm3 in the filamentary structures.
High densities  (\n0 =100-80 \cm3) are found in the Arched Filaments,  the maximum values
(\n0=200 \cm3) in E91 position C, and in C96 positions 7 and 11 (280 and 200 \cm3, respectively).

The magnetic field  ranges from 5. 10$^{-6}$ gauss  in  S07 positions 1 and 2, characteristic of the ISM,
  increasing smoothly  to
$>$ 5 10$^{-5}$ gauss beyond the Bubble, up to a maximum of 8 10$^{-5}$ gauss. These values are  about the same 
as found in the crossing
strip E2-W1 in the Arched Filaments. Beyond the Arched Filaments, \B0 regains the ISM values.
Our results confirm LaRosa et al. (2005) predictions of the magnetic field strength.

The maximum temperature of the stars  are higher in the Quintuplet Cluster ($\sim$ 38000 K)
than in the Arches Cluster ($\sim$ 27000 K). There are stars at temperatures of $\sim$ 35000 K  
in the southern  ISM
 and of $\sim$ 39000 K in the  northern one,  above 0.1 degree.
The ionization parameter in relatively low ($<$ 0.01 )  reaching
 a maximum of $>$0.01 near the Arches Cluster.  This indicates that the observed positions 30-35
are closer to the stars. In the E2-W1 strip, U is rather low, diluted by the distance
from the ionization source, most probably the Arches Cluster.

The depletion from the gaseous phase of Si is ubiquitous, indicating the presence of
silicate dust throughout all the region, while a  large quantity of iron rich  grains 
is present in the region of the Arched Filaments.

Comparing the relative abundances for positions 29-34, S07 find the average Ne/H=1.63$\pm$0.08 10$^{-4}$, 
S/H 1.16$\pm$0.06 10$^{-5}$,
 while we find that Ne/H $\sim$ 10$^{-4}$  satisfactorily fits all positions and S/H fluctuates
between 6.3 10$^{-6}$ and 1.6 10$^{-5}$.
S07 find Fe/H $\sim$ 1.3 10$^{-6}$ in the Arched Filament and $\sim$ 8.8 10$^{-6}$ in the Bubble,
in agreement with our results :Fe/H  $\sim$ 10$^{-5}$ in the Bubble and 10$^{-6}$ in the Arched Filaments
(Fig. 3c).

The continuum SED  between 33 and 158 \mum in all the observed regions  indicate that a component
of dust heated to temperatures of $\sim$ 100-200 K must be present. The  dust grains coupled to gas
within the emitting clouds  cannot reach those high temperatures by using the input parameters
which are constrained by the fit of the line spectra. Higher ionization parameters and  small
grains characterise this dust. We suggest that hot  dust is located closer to the  stars
than the emitting gaseous clumps.
The temperature of the stars is not high enough to destroy the grains by evaporation,
and  the shock velocity cannot  disrupt them totally by sputtering.
In the Arched Filaments, we find a dust-to-gas ratio  $\sim$ 0.4 by mass. 

The data are  insufficient
 to show absorption and emission bands from the grains  or
 to constrain the dust-to-gas ratios in
the different regions.
PAHs can be present in  some Arched filament region positions,
 leading to a strong absorption of the photoionizing flux from the stars.

The radio emission  seems thermal bremsstrahlung in all the positions observed in  the Arched Filaments,
however a synchrotron radiation component is not excluded.
More data should confirm and improve the modelling presented in this paper.

\section*{Acknowledgements}
I am grateful to J. P. Simpson, E. F. Erickson, and S. W. Colgan, for allowing me to reproduce their 
figures, and to an anonymous referee for many important comments. I thank R. Angeloni for helpful advise.

\end{document}